\documentclass[aps,pra,twocolumn,floats]{revtex4-2}
\usepackage{amsmath}
\usepackage{amssymb}
\usepackage{bm}
\usepackage{graphicx}
\usepackage{xcolor}
\usepackage{hyperref}

\newcommand{\ryvOm}{\boldsymbol{\Omega}}
\newcommand{\rygn}{\boldsymbol{\nabla}}
\newcommand{\rygnp}{\boldsymbol{\nabla}_{\mathbf p}}

\newcommand{\rytodo}[1]{}

\begin{document}

\title{\textbf{Ray propagation of circularly polarized light in an inhomogeneous medium:
the optical Magnus effect from Maxwell's equations}}

\author{Igor Kuzmenko$^{1,2}$, Y. B. Band$^{1,2,3}$}

\affiliation{
  $^1$Department of Physics,
  Ben-Gurion University of the Negev,
  Beer-Sheva 84105, Israel
  \\
  $^2$Department of Chemistry,
  Ben-Gurion University of the Negev,
  Beer-Sheva 84105, Israel
  \\
  $^3$The Ilse Katz Center for Nano-Science,
  Ben-Gurion University of the Negev,
  Beer-Sheva 84105, Israel
  }

\date{\today}

\begin{abstract}
We derive the ray propagation of circularly polarized light in an inhomogeneous,
isotropic and lossless medium, including the first correction to geometrical
optics, the optical Magnus (spin Hall) effect. Starting from Maxwell's equations and
the eikonal expansion to first order in $1/k_0$, we show that the polarization is
parallel transported along the ray, identify the Berry connection and curvature of the
helicity states, and obtain a ray action in which the Berry phase appears as a
correction to the optical path length. This action leads to a Hamiltonian in which the
refractive index is evaluated at a position shifted by the Berry connection,
$\mathcal H_\sigma(\mathbf r,\mathbf p)=|\mathbf p|-n(\mathbf r+\mathbf A_\sigma(\mathbf p)/k_0)
\simeq|\mathbf p|-n(\mathbf r)-k_0^{-1}\boldsymbol\nabla n\cdot\mathbf A_\sigma(\mathbf p)$.
In terms of the physical position, the equations of motion contain the Berry curvature as
an anomalous velocity. We give the equivalent non-canonical bracket form, the gauge
transformation, and the equivalent time-parametrized formulation with the Hamiltonian
$\hbar\omega=\hbar c|\mathbf k|/n$, and we state the approximations involved. The resulting ray equations are compared with analytic and numerical results. A planar interface, the helical ray in a cylindrical medium and the exact conservation of angular momentum provide checks. The formulation is set up for numerical propagation of rays in a medium with a spatially varying refractive index.
\end{abstract}

\maketitle

\section{Introduction}\label{sec:intro}

In geometrical optics, light in a medium of refractive index $n(\mathbf r)$ follows
the rays of a Hamiltonian system. The first correction in the wavelength depends on the
helicity: right and left circularly polarized waves follow slightly different
trajectories. This is the optical Magnus effect, or spin Hall effect of light
\cite{Liberman1992,BliokhPRE2004,BliokhPLA2004,Onoda2004,Bliokh2008}. In the
language of semiclassical dynamics it is an anomalous-velocity term generated by
the Berry curvature in momentum space \cite{Sundaram1999}.

The effect is described in the literature in two equivalent forms. In the first
\cite{BliokhPRE2004,Bliokh2008}, the equations of motion contain the Berry curvature
explicitly. In the second, the helicity is built into the Hamiltonian through the Berry
connection, as a shift of the position at which the refractive index is evaluated.
In this paper we derive the helicity-dependent ray equations from Maxwell's equations,
show how the two forms are related, make the gauge dependence explicit, state the
approximations involved, and set up the equations for the numerical propagation of rays
in a medium with a spatially varying refractive index.
The effect was predicted by Liberman and Zel'dovich \cite{Liberman1992}, derived for a smoothly inhomogeneous medium by Bliokh and Bliokh \cite{BliokhPRE2004,BliokhPLA2004}, and described as a Hall effect of light, with the semiclassical equations of motion of a wave packet, by Onoda, Murakami and Nagaosa \cite{Onoda2004}; reviews are given in Refs.~\cite{Bliokh2009,Bliokh2015}. The helicity-dependent deflection was observed for helical rays in a glass cylinder \cite{Bliokh2008}, where the displacement, of the order of the wavelength per turn of the direction of propagation, accumulates over many turns. The associated Berry phase, the modern form of the Rytov-Vladimirskii law of rotation of the polarization plane along a curved ray \cite{Rytov1938,Vladimirskii1941,Berry1987}, was measured in helically coiled optical fibers \cite{TomitaChiao1986}. At a sharp interface the corresponding effect is the Imbert-Fedorov shift \cite{BliokhIF2006,HostenKwiat2008}. The same ray equations describe light in gravitational fields \cite{Gosselin2007,Oancea2020,Yamamoto2017} and, for light carrying orbital angular momentum, in an optical waveguide \cite{Qiu2024}; numerical integrations of the helicity-dependent ray equations have been reported for the Schwarzschild spacetime \cite{Oancea2020} and for such a waveguide \cite{Qiu2024}. We are not aware of a numerical propagation for a general three-dimensional gradient-index profile.

The new elements of this paper are the following. (i) The parallel-transport law of the polarization and the helicity-dependent ray equations are derived from the first-order Maxwell equations for a medium in which both the permittivity and the permeability vary; the polarization transport and the ray equations depend only on the refractive index, while the impedance enters only the amplitude. (ii) The relation between the shifted-argument (Berry-connection) Hamiltonian and the Berry-curvature equations of motion is made explicit, including the gauge dependence of the canonical coordinate and the identification of the gauge-invariant physical position of the ray. (iii) The exact conservation law of the total angular momentum is given in the form used for numerical integration, together with a set of analytic tests. (iv) The ray equations are integrated numerically for a three-dimensional gradient-index profile. The equivalent formulations (Lagrangian of the ray path, non-canonical brackets, and frequency as the Hamiltonian) and the detailed comparison with Bliokh's review \cite{Bliokh2009} are collected in the appendices.

The paper is organized as follows. In Sec.~\ref{sec:maxwell} we write the eikonal
expansion of Maxwell's equations and show that the polarization of the first-order field
is parallel transported along the ray. Section~\ref{sec:berry} defines the circular
polarization basis, the Berry phase, the Berry connection and the Berry curvature.
Section~\ref{sec:action} gives the ray action and the Hamiltonian including the Magnus term, derives the equations of motion in terms of the physical
position, and gives the conservation law of the total angular momentum. Section~\ref{sec:validity} estimates the size of the effect and discusses the validity of the description, Sec.~\ref{sec:interface} treats a planar interface as a check, and Sec.~\ref{sec:numerical} presents the numerical propagation and the tests of the numerical calculation. Section~\ref{sec:summary} summarizes the results. Appendix~\ref{app:transport} gives the details of the first-order solvability condition, Appendix~\ref{app:frenet} the explicit Berry connections and the Frenet-Serret construction, Appendix~\ref{app:centroid} the centroid of a wave packet, Appendix~\ref{app:formulations} the equivalent formulations of the ray dynamics, and Appendix~\ref{app:literature} the comparison with Bliokh's review and the analytic results for a helical ray.

\section{Maxwell's equations and the eikonal expansion}\label{sec:maxwell}

The equations of this section are written in SI units. The ray equations that follow involve only the dimensionless optical momentum $\mathbf p$ and the vacuum wave number $k_0=\omega/c$, and are independent of the system of units.

\subsection{Maxwell's equations}\label{subsec:maxwell}

Consider a monochromatic electromagnetic wave with time dependence $e^{-i\omega t}$
propagating through an inhomogeneous, isotropic and lossless medium with
position-dependent permittivity $\epsilon(\mathbf r)$ and permeability $\mu(\mathbf r)$.
The Faraday and Amp\`ere equations are
\begin{align}
  \label{eq:Faraday}
  \rygn\times\boldsymbol{\mathcal E}(\mathbf r) &= i\omega\,\mu(\mathbf r)\,\boldsymbol{\mathcal H}(\mathbf r),
  \\
  \label{eq:Ampere}
  \rygn\times\boldsymbol{\mathcal H}(\mathbf r) &= -i\omega\,\epsilon(\mathbf r)\,\boldsymbol{\mathcal E}(\mathbf r).
\end{align}
The local refractive index and the local impedance are
\begin{align}
n(\mathbf r) &=
\sqrt{\frac{\epsilon(\mathbf r)\mu(\mathbf r)}{\epsilon_0\mu_0}} =
c\sqrt{\epsilon(\mathbf r)\mu(\mathbf r)} ,
\label{eq:refractive-index}\\
\mathcal Z(\mathbf r) &= \sqrt{\frac{\mu(\mathbf r)}{\epsilon(\mathbf r)}} ,
\label{eq:impedance}
\end{align}
where $c=1/\sqrt{\epsilon_0\mu_0}$ is the speed of light in a vacuum. The vacuum wave
number is $k_0=2\pi/\lambda_0=\omega/c$. The medium is assumed to vary slowly on the
scale of the vacuum wavelength $\lambda_0$.

\subsection{Eikonal expansion}\label{subsec:Eikonal}

To study the short-wavelength limit ($k_0\to\infty$), using the standard method of geometrical optics \cite{KravtsovOrlov1990}, we separate the rapidly varying
phase (the eikonal $\psi(\mathbf r)$) from slowly varying complex vector amplitudes
$\mathbf E(\mathbf r)$ and $\mathbf H(\mathbf r)$:
\begin{align}
\label{eq:eikonal-E-field}
\boldsymbol{\mathcal E}(\mathbf r) &= \mathbf E(\mathbf r)\,e^{ik_0\psi(\mathbf r)} ,
\\
\label{eq:eikonal-H-field}
\boldsymbol{\mathcal H}(\mathbf r) &= \mathbf H(\mathbf r)\,e^{ik_0\psi(\mathbf r)} .
\end{align}
We define the optical momentum
\begin{equation}
\mathbf p=\rygn\psi .
\label{eq:p-vector}
\end{equation}
Since $\rygn\times(\mathbf V e^{ik_0\psi})=(\rygn\times\mathbf V+ik_0\,\mathbf p\times\mathbf V)e^{ik_0\psi}$ for any vector amplitude $\mathbf V$,
Eqs.~(\ref{eq:Faraday}) and (\ref{eq:Ampere}) become
\begin{align}
\rygn\times\mathbf E+ik_0\,\mathbf p\times\mathbf E &= ik_0\,c\,\mu\,\mathbf H ,
\label{eq:Faraday-eikonal}\\
\rygn\times\mathbf H+ik_0\,\mathbf p\times\mathbf H &= -ik_0\,c\,\epsilon\,\mathbf E .
\label{eq:Ampere-eikonal}
\end{align}
We expand the amplitudes in powers of $1/(ik_0)$,
\begin{align}
\mathbf E &= \sum_{l=0}^{\infty}\frac{1}{(ik_0)^l}\,\mathbf E_l ,
&
\mathbf H &= \sum_{l=0}^{\infty}\frac{1}{(ik_0)^l}\,\mathbf H_l .
\label{eq:eh-series}
\end{align}
Inserting the series into Eqs.~(\ref{eq:Faraday-eikonal}) and (\ref{eq:Ampere-eikonal}) and
collecting the coefficients of each power of $ik_0$, we obtain the zeroth-order equations
\begin{align}
  \mathbf p\times\mathbf E_0-c\mu\,\mathbf H_0 &= 0 ,
  \label{eq:Faraday-eikonal-0-th}\\
  \mathbf p\times\mathbf H_0+c\epsilon\,\mathbf E_0 &= 0 ,
  \label{eq:Ampere-eikonal-0-th}
\end{align}
and, for $l\ge1$,
\begin{align}
  \rygn\times\mathbf E_{l-1}+\mathbf p\times\mathbf E_l-c\mu\,\mathbf H_l &= 0 ,
  \label{eq:Faraday-eikonal-l-th}\\
  \rygn\times\mathbf H_{l-1}+\mathbf p\times\mathbf H_l+c\epsilon\,\mathbf E_l &= 0 .
  \label{eq:Ampere-eikonal-l-th}
\end{align}
Section~\ref{subsec:zero-order} solves the zeroth-order equations; Sec.~\ref{subsec:first-order}
treats the case $l=1$ and the condition for the existence of the first-order fields.

\subsection{Zeroth order}\label{subsec:zero-order}

Eliminating $\mathbf H_0$ between Eqs.~(\ref{eq:Faraday-eikonal-0-th}) and
(\ref{eq:Ampere-eikonal-0-th}) gives
\begin{equation}
\mathbf p\times[\mathbf p\times\mathbf E_0]+n^2\mathbf E_0=0 ,
\label{eq:for-e_0}
\end{equation}
with $n$ given by Eq.~(\ref{eq:refractive-index}). Equation~(\ref{eq:for-e_0}) has nontrivial
solutions if and only if
\begin{equation}
\det\big[p_jp_{j'}+(n^2-p^2)\delta_{jj'}\big]=n^2\,(p^2-n^2)^2=0 ,
\label{eq:det-M-vanishes}
\end{equation}
(the matrix has the eigenvalue $n^2$ for the eigenvector $\mathbf p$ and the doubly degenerate eigenvalue $n^2-p^2$ for vectors transverse to $\mathbf p$) which gives the eikonal equation
\begin{equation}
p=|\mathbf p|=n(\mathbf r),
\label{eq:eikonal-equation}
\end{equation}
since $p$ and $n$ are real and positive. Equations~(\ref{eq:Faraday-eikonal-0-th}) and
(\ref{eq:Ampere-eikonal-0-th}) then give $\mathbf p\cdot\mathbf E_0=\mathbf p\cdot\mathbf H_0=0$
and
\begin{equation}
\mathbf H_0=\frac{1}{c\mu}\,\mathbf p\times\mathbf E_0 ,
\label{eq:H0-vs-E0}
\end{equation}
that is, the zeroth-order fields are transverse. In the sense of the bilinear scalar
product, $\{\mathbf E_0,\mathbf H_0,\mathbf p\}$ are mutually orthogonal. The Poynting
vector points along $\mathbf p$, so that the rays are the curves $\mathbf r(\ell)$ with
tangent $\hat{\mathbf p}=\mathbf p/p=d\mathbf r/d\ell$, where $\ell$ is the arclength.
Equation~(\ref{eq:eikonal-equation}) implies $\psi(\mathbf r(\ell))=\int_0^\ell n\,d\ell'$ and
the ray equation $d(n\hat{\mathbf p})/d\ell=\rygn n$, which is equivalent to
\begin{equation}
\frac{d\mathbf r}{d\ell}=\hat{\mathbf p},\qquad
\frac{d\mathbf p}{d\ell}=\rygn n .
\label{eq:ray0}
\end{equation}
Here $\mathbf r$ is a point of space, and to this order the curve $\mathbf r(\ell)$ is the position of the ray at arclength $\ell$. At first order in $1/k_0$ two different curves appear, the position $\mathbf R(\ell)$ of the ray and the canonical coordinate $\mathbf r(\ell)$; the reason why $\mathbf R$ is the position of the light is given in Sec.~\ref{subsec:why-R}.

\subsection{First order and the solvability condition}\label{subsec:first-order}

For $l=1$, Eqs.~(\ref{eq:Faraday-eikonal-l-th}) and (\ref{eq:Ampere-eikonal-l-th}) read
\begin{align}
\mathbf p\times\mathbf E_1-c\mu\,\mathbf H_1 &= -\rygn\times\mathbf E_0 ,
\label{eq:Faraday-eikonal-1-th}\\
\mathbf p\times\mathbf H_1+c\epsilon\,\mathbf E_1 &= -\rygn\times\mathbf H_0 .
\label{eq:Ampere-eikonal-1-th}
\end{align}
This is an inhomogeneous linear system whose homogeneous part has vanishing
determinant, so it has a solution only if the inhomogeneous terms satisfy a condition. Solving Eq.~(\ref{eq:Faraday-eikonal-1-th}) for $\mathbf H_1$,
\begin{equation}
\mathbf H_1=\frac{1}{c\mu}\big[\mathbf p\times\mathbf E_1+\rygn\times\mathbf E_0\big] ,
\end{equation}
and inserting the result into Eq.~(\ref{eq:Ampere-eikonal-1-th}), multiplied by $c\mu$, we find
\begin{equation}
\mathbf p\times[\mathbf p\times\mathbf E_1]+n^2\mathbf E_1=\boldsymbol{\mathcal C} ,
\label{eq:matrix-append}
\end{equation}
where
\begin{equation}
\boldsymbol{\mathcal C}=-c\mu\,\rygn\times\mathbf H_0-\mathbf p\times(\rygn\times\mathbf E_0) .
\label{eq:B-vector}
\end{equation}
On the surface $p=n$, Eq.~(\ref{eq:matrix-append}) is
\begin{equation}
\mathcal M\,\mathbf E_1=\mathbf p\,(\mathbf p\cdot\mathbf E_1)=\boldsymbol{\mathcal C} ,
\qquad
\mathcal M_{jj'}=p_jp_{j'}+(n^2-p^2)\delta_{jj'} ,
\label{eq:M-E1}
\end{equation}
because $\mathcal M$ has rank one when $p=n$. Equation~(\ref{eq:M-E1}) can be solved if and only if
\begin{equation}
\boldsymbol{\mathcal C}\ \text{is collinear with}\ \mathbf p ,
\label{eq:solvability}
\end{equation}
i.e., the components of $\boldsymbol{\mathcal C}$ transverse to $\mathbf p$ vanish. (Equivalently, with the notation $\Delta=\det\mathcal M$ and with $\Delta_{j''}$ the determinant of $\mathcal M$ with its $j''$th column replaced by $\boldsymbol{\mathcal C}$, Cramer's rule gives $\Delta_{j''}/\Delta=[n^2C_{j''}-p_{j''}(\mathbf p\cdot \boldsymbol{\mathcal C})]/[n^2(n^2-p^2)]$, whose limit $p\to n$ is finite only if the numerator vanishes at $p=n$, which is Eq.~(\ref{eq:solvability}).)

We evaluate $\boldsymbol{\mathcal C}$ for fields that vary along the ray,
\begin{equation}
\rygn\times\mathbf X=\hat{\mathbf p}\times\frac{d\mathbf X}{d\ell} ,
\label{eq:curl-ray}
\end{equation}
which neglects the derivatives transverse to the ray (the spreading of the ray bundle).  With this approximation both terms of $\boldsymbol{\mathcal C}$ are orthogonal to $\hat{\mathbf p}$, so that $\mathbf p\cdot\boldsymbol{\mathcal C}=0$, and the condition (\ref{eq:solvability}) becomes $\boldsymbol{\mathcal C}=0$.

To find $\boldsymbol{\mathcal C}$, introduce the normal $\hat{\boldsymbol\nu}$ and the
binormal $\hat{\boldsymbol\beta}=\hat{\mathbf p}\times\hat{\boldsymbol\nu}$ of the ray, so that
$\{\hat{\mathbf p},\hat{\boldsymbol\nu},\hat{\boldsymbol\beta}\}$ is a right-handed orthonormal
Frenet-Serret frame \cite{Tu2017},
\begin{equation}
\frac{d\hat{\mathbf p}}{d\ell}=\mathcal K\,\hat{\boldsymbol\nu},
\quad
\frac{d\hat{\boldsymbol\nu}}{d\ell}=-\mathcal K\,\hat{\mathbf p}+\mathcal T\,\hat{\boldsymbol\beta},
\quad
\frac{d\hat{\boldsymbol\beta}}{d\ell}=-\mathcal T\,\hat{\boldsymbol\nu},
\label{eq:Frenet-Serret-formulas}
\end{equation}
where $\mathcal K=|d\hat{\mathbf p}/d\ell|$ is the curvature and
$\mathcal T=-\hat{\boldsymbol\nu}\cdot d\hat{\boldsymbol\beta}/d\ell$ the torsion of the ray.
The transverse field is
\begin{equation}
\mathbf E_0=E_\nu\,\hat{\boldsymbol\nu}+E_\beta\,\hat{\boldsymbol\beta} ,
\label{eq:e_0-general}
\end{equation}
with complex components $E_\nu$ and $E_\beta$, and Eq.~(\ref{eq:H0-vs-E0}) gives
\begin{equation}
\mathbf H_0=\mathcal Y\,\big[-E_\beta\,\hat{\boldsymbol\nu}+E_\nu\,\hat{\boldsymbol\beta}\big],
\qquad \mathcal Y=\frac{1}{\mathcal Z}=\sqrt{\frac{\epsilon}{\mu}} .
\label{eq:h_0-general}
\end{equation}
Evaluating $\boldsymbol{\mathcal C}$ with Eqs.~(\ref{eq:curl-ray})--(\ref{eq:h_0-general}) (Appendix~\ref{app:transport}) gives $\boldsymbol{\mathcal C}=2n[D_\nu\hat{\boldsymbol\nu}+D_\beta\hat{\boldsymbol\beta}]+n(d\ln\mathcal Y/d\ell)\mathbf E_0$, where $D_\nu$ and $D_\beta$ are the covariant derivatives of the components of $\mathbf E_0$ along the ray [Eq.~(\ref{eq:D-def})].
The condition $\boldsymbol{\mathcal C}=0$ therefore reads
\begin{equation}
\Big(\frac{d\mathbf E_0}{d\ell}\Big)_\perp=-\frac12\,\frac{d\ln\mathcal Y}{d\ell}\,\mathbf E_0 ,
\label{eq:parallel-transport}
\end{equation}
where $(\cdot)_\perp$ is the part orthogonal to $\hat{\mathbf p}$. For a constant impedance this is the
statement that the polarization vector $\mathbf E_0$ is \emph{parallel transported} along
the ray, $D_\nu=D_\beta=0$ (Rytov-Vladimirskii law). Components of $\mathbf E_0$ along $\hat{\boldsymbol\nu}$ and
$\hat{\boldsymbol\beta}$ rotate into each other at the rate $\mathcal T$, so that the polarization
plane of linearly polarized light rotates by the angle $\int\mathcal T\,d\ell$. The impedance factor in
Eq.~(\ref{eq:parallel-transport}) gives the amplitude $\propto {\mathcal Z}^{1/2}$, i.e., it keeps $|\mathbf E_0|^2/{\mathcal Z}$, and hence the energy flux along a ray tube of fixed cross section, constant.

\section{Helicity states, Berry phase and Berry curvature}\label{sec:berry}

\subsection{Circular polarization basis and uncoupled helicity states}\label{subsec:circular}

We use the local circular polarization vectors
\begin{equation}
\hat{\mathbf e}_\sigma=\frac{1}{\sqrt2}\big[\hat{\boldsymbol\nu}+i\sigma\,\hat{\boldsymbol\beta}\big],
\qquad \sigma=\pm1 .
\label{eq:e_sigma}
\end{equation}
With the time dependence $e^{-i\omega t}$, the real field of $\hat{\mathbf e}_\sigma$ is
proportional to $\hat{\boldsymbol\nu}\cos\omega t+\sigma\hat{\boldsymbol\beta}\sin\omega t$.
Because $\hat{\boldsymbol\nu}\times\hat{\boldsymbol\beta}=\hat{\mathbf p}$, for $\sigma=+1$ the field rotates
counterclockwise as seen by an observer facing the source, so that $\sigma=+1$ is positive helicity (spin along $\hat{\mathbf p}$). This is left circular polarization in the
convention of classical optics, and right circular polarization in the convention of the IEEE.
In this paper $\sigma=+1$ and $\sigma=-1$ denote positive and negative helicity.

For circular polarization we write $E_\nu=\mathcal E_\sigma/\sqrt2$ and $E_\beta=i\sigma\mathcal E_\sigma/\sqrt2$, so that
\begin{equation}
\mathbf E_0=\mathcal E_\sigma\,\hat{\mathbf e}_\sigma ,
\qquad
\mathbf H_0=-i\sigma\,\mathcal Y\,\mathcal E_\sigma\,\hat{\mathbf e}_\sigma ,
\label{eq:E0-H0-circular}
\end{equation}
where we used Eq.~(\ref{eq:h_0-general}). Then $D_\beta=i\sigma D_\nu$, and the two components of the
condition~(\ref{eq:parallel-transport}) reduce to a single equation,
\begin{equation}
\frac{d\mathcal E_\sigma}{d\ell}-i\sigma\mathcal T\,\mathcal E_\sigma+\frac12\frac{d\ln\mathcal Y}{d\ell}\,\mathcal E_\sigma=0 ,
\label{eq:for-E_sigma}
\end{equation}
with the solution
\begin{equation}
\mathcal E_\sigma(\ell)=\Big[\frac{\mathcal Y(0)}{\mathcal Y(\ell)}\Big]^{1/2}
\exp\Big(i\sigma\int_0^\ell\mathcal T(\ell')\,d\ell'\Big)\,\mathcal E_\sigma(0) .
\label{eq:E_sigma-solution}
\end{equation}
The two helicity states are therefore uncoupled: the transverse part of $d\hat{\mathbf e}_\sigma/d\ell$ is proportional to
$\hat{\mathbf e}_\sigma$ itself [Eq.~(\ref{eq:de-frenet}) of Appendix~\ref{app:frenet}], and the
helicity is not changed by propagation to this order. A linearly polarized wave is a superposition of the two helicity
states, whose phases $\pm\int\mathcal T\,d\ell$ differ, which is the rotation of the polarization plane
by $\int\mathcal T\,d\ell$.

\subsection{Berry phase and Berry connection}\label{subsec:Berry-phase-connection}

The solution (\ref{eq:E_sigma-solution}) contains the phase
\begin{equation}
\varphi_\sigma(\ell)=\sigma\int_0^\ell\mathcal T(\ell')\,d\ell' ,
\label{eq:Berry-phase-frenet}
\end{equation}
which is defined with respect to the Frenet basis of the ray. The same phase can be expressed through the
momentum $\mathbf p(\ell)$. Let $\hat{\mathbf e}_\sigma(\mathbf p)$ be a smooth choice (a gauge) of the circular
polarization vector as a function of $\mathbf p$ only, and define the Berry connection
\begin{equation}
\mathbf A_\sigma(\mathbf p)=i\,\hat{\mathbf e}_\sigma^{\,*}\cdot\rygnp\,\hat{\mathbf e}_\sigma
=i\,\langle\hat{\mathbf e}_\sigma|\rygnp|\hat{\mathbf e}_\sigma\rangle ,
\label{eq:A_sigma}
\end{equation}
which is real because $\hat{\mathbf e}_\sigma^{\,*}\cdot\hat{\mathbf e}_\sigma=1$. Writing
$\mathbf E_0=\mathcal E_\sigma\,\hat{\mathbf e}_\sigma(\mathbf p(\ell))$ and projecting
Eq.~(\ref{eq:parallel-transport}) on $\hat{\mathbf e}_\sigma^{\,*}$ gives
\begin{align}
&\frac{d\mathcal E_\sigma}{d\ell}+\langle\hat{\mathbf e}_\sigma|\tfrac{d}{d\ell}\hat{\mathbf e}_\sigma\rangle\mathcal E_\sigma
+\frac12\frac{d\ln\mathcal Y}{d\ell}\,\mathcal E_\sigma=0 ,
\nonumber\\
&\langle\hat{\mathbf e}_\sigma|\tfrac{d}{d\ell}\hat{\mathbf e}_\sigma\rangle=-i\,\mathbf A_\sigma\cdot\frac{d\mathbf p}{d\ell} ,
\end{align}
so that
\begin{gather}
\mathcal E_\sigma(\ell)=\Big[\frac{\mathcal Y(0)}{\mathcal Y(\ell)}\Big]^{1/2}e^{i\varphi_\sigma(\ell)}\mathcal E_\sigma(0),
\\
\varphi_\sigma(\ell)=\int\mathbf A_\sigma(\mathbf p)\cdot d\mathbf p .
\label{eq:Berry-phase}
\end{gather}
This is the Berry phase of the helicity state, in the gauge in which $\hat{\mathbf e}_\sigma$ is defined. In the
Frenet gauge, $\hat{\mathbf e}_\sigma=(\hat{\boldsymbol\nu}+i\sigma\hat{\boldsymbol\beta})/\sqrt2$, one has $\mathbf A_\sigma\cdot d\mathbf p/d\ell=\sigma\mathcal T$, which reproduces
Eq.~(\ref{eq:Berry-phase-frenet}) (Appendix~\ref{app:frenet}). For a different choice of the gauge the phase of an open path changes by a gauge term; see Eq.~(\ref{eq:gauge-phase}) below and Appendix~\ref{app:frenet}. The parallel-transport equation (\ref{eq:Berry-phase}) is the same as the evolution of the state of a quantum system with adiabatically varying parameters ${\boldsymbol {\mathcal R}}$, where the Berry connection is
$\mathbf A_{m}=i\langle u_m|\nabla_{{\boldsymbol {\mathcal R}}}|u_m\rangle$ ($u_m$ the instantaneous eigenstate); here the parameters are the components of $\mathbf p$ and
the ``time'' is the arclength. The fields are
\begin{align}
\boldsymbol{\mathcal E}(\mathbf r)&=\mathcal E_{\rm r}\,\hat{\mathbf e}_\sigma(\mathbf r)\,e^{ik_0\psi(\mathbf r)+i\varphi_\sigma(\mathbf r)} ,
\label{eq:eikonal-E-field_sigma}\\
\boldsymbol{\mathcal H}(\mathbf r)&=-i\sigma\,\mathcal E_{\rm r}\,\mathcal Y(\mathbf r)\,\hat{\mathbf e}_\sigma(\mathbf r)\,e^{ik_0\psi(\mathbf r)+i\varphi_\sigma(\mathbf r)} ,
\label{eq:eikonal-H-field_sigma}
\end{align}
where $\mathcal E_{\rm r}(\mathbf r)\propto\mathcal Y^{-1/2}$ is the real amplitude, and $\psi(\mathbf r(\ell))=\int_0^\ell n\,d\ell'$.

\subsection{Berry curvature and gauge transformations}\label{subsec:Berry-curvature}

The Berry curvature of the helicity state $\sigma$ is gauge invariant and is the same for every smooth choice of $\hat{\mathbf e}_\sigma(\mathbf p)$; with the explicit connections of Appendix~\ref{app:frenet},
\begin{equation}
\ryvOm_\sigma(\mathbf p)=\rygnp\times\mathbf A_\sigma(\mathbf p)=-\sigma\,\frac{\mathbf p}{p^{3}} ,
\label{eq:Berry-curvature}
\end{equation}
the field of a magnetic monopole of strength $-\sigma$ at $\mathbf p=0$.

Under a gauge transformation $\hat{\mathbf e}_\sigma\to e^{i\sigma\chi(\mathbf p)}\hat{\mathbf e}_\sigma$ (so that
$\hat{\mathbf e}_\sigma$ is multiplied by a phase), the connection and the Berry phase of an open path change as
\begin{equation}
\mathbf A_\sigma\to\mathbf A_\sigma-\sigma\rygnp\chi ,
\qquad
\varphi_\sigma\to\varphi_\sigma-\sigma\,\Delta\chi ,
\label{eq:gauge-phase}
\end{equation}
where $\Delta\chi$ is the change of $\chi$ along the path. Only the curvature, and the phase of a closed loop in
$\mathbf p$ space, which equals the flux of $\ryvOm_\sigma$ through the loop, i.e., $-\sigma$ times the solid angle
enclosed by $\hat{\mathbf p}$ (modulo $2\pi$), are gauge invariant. The equations of motion
derived in Sec.~\ref{sec:action} contain only $\ryvOm_\sigma$.

\section{Ray action, Hamiltonian and equations of motion}\label{sec:action}

\subsection{Optical path length with the Berry phase}\label{subsec:path}

The phase of the field in Eq.~(\ref{eq:eikonal-E-field_sigma}) is $k_0\psi+\varphi_\sigma$. Dividing by $k_0$, the
optical path length of a ray from $\mathbf r_i$ to $\mathbf r_f$ is
\begin{equation}
\mathfrak L_\sigma=\int_0^\ell\Big[n(\mathbf r(\ell'))
+\frac{1}{k_0}\,\mathbf A_\sigma(\mathbf p)\cdot\frac{d\mathbf p}{d\ell'}\Big]d\ell' ,
\label{eq:path}
\end{equation}
where the first term is the usual optical path and the second is the spin-dependent Berry phase accumulated along the
trajectory, divided by $k_0$. The rays are the stationary paths of $\mathfrak L_\sigma$ (Fermat's principle with the Berry phase),
$\delta\mathfrak L_\sigma=0$. The Berry term is a correction of order $1/k_0$ to the usual optical path.

\subsection{Phase-space action}\label{subsec:phase-space}

Equation~(\ref{eq:path}) can be written as a phase-space action in which the momentum $\mathbf p$ is an independent
variable and the coordinate is the position $\mathbf R(\ell)$ of the ray\ (this identification is justified in Sec.~\ref{subsec:why-R}),
\begin{equation}
S=\int d\ell\Big[\mathbf p\cdot\frac{d\mathbf R}{d\ell}+\frac{1}{k_0}\,\mathbf A_\sigma(\mathbf p)\cdot\frac{d\mathbf p}{d\ell}
-\big(|\mathbf p|-n(\mathbf R)\big)\Big] .
\label{eq:phase-space-action}
\end{equation}
(Eliminating $\mathbf p$ with $d\mathbf R/d\ell=\hat{\mathbf p}$ and $|\mathbf p|=n$ gives back Eq.~(\ref{eq:path}).)
Varying $\mathbf p$ and $\mathbf R$, and using the identity
\begin{equation}
\frac{\partial A_j}{\partial p_i}-\frac{\partial A_i}{\partial p_j}=\varepsilon_{ijl}\,\Omega_l ,
\label{eq:identity}
\end{equation}
which follows from $\Omega_l=\varepsilon_{lmn}\partial_{p_m}A_n$ and
$\varepsilon_{ijl} \, \varepsilon_{lmn}=\delta_{im}\delta_{jn}-\delta_{in}\delta_{jm}$ (here $\mathbf A=\mathbf A_\sigma$ and
$\ryvOm=\ryvOm_\sigma$), one obtains the closed system
\begin{align}
\frac{d\mathbf R}{d\ell}&=\hat{\mathbf p}-\frac{1}{k_0}\,\frac{d\mathbf p}{d\ell}\times\ryvOm_\sigma(\mathbf p) ,
\label{eq:dR-dl-general}\\
\frac{d\mathbf p}{d\ell}&=\rygn n(\mathbf R) .
\label{eq:dp-dl-general}
\end{align}
Inserting Eq.~(\ref{eq:dp-dl-general}) into Eq.~(\ref{eq:dR-dl-general}) and using Eq.~(\ref{eq:Berry-curvature}),
\begin{align}
\frac{d\mathbf R}{d\ell}&=\hat{\mathbf p}-\frac{1}{k_0}\,\rygn n(\mathbf R)\times\ryvOm_\sigma(\mathbf p)
= \frac{\mathbf p}{|\mathbf p|}+\frac{\sigma}{k_0}\,\rygn n(\mathbf R)\times\frac{\mathbf p}{p^{3}} ,
\label{eq:HamEqr_cor}\\
\frac{d\mathbf p}{d\ell}&=\rygn n(\mathbf R) .
\label{eq:HamEqp_cor}
\end{align}
These are the helicity-dependent ray equations in terms of the physical position $\mathbf R$ and the optical momentum
$\mathbf p$. The quantity $|\mathbf p|-n(\mathbf R)$ is conserved, since
$\frac{d}{d\ell}(|\mathbf p|-n(\mathbf R))=-\rygn n\cdot(\rygn n\times\ryvOm_\sigma)/k_0=0$, so that $|\mathbf p|=n(\mathbf R)$ holds along
the whole ray if it holds initially. On this surface the second term of Eq.~(\ref{eq:HamEqr_cor}) is
\begin{equation}
\frac{\sigma}{k_0n^{2}}\,\rygn n\times\hat{\mathbf p} .
\label{eq:Magnus-term}
\end{equation}
It is the optical Magnus term: it is transverse to both $\rygn n$ and $\hat{\mathbf p}$, it is proportional to the vacuum wavelength $\lambda_0$, and it has opposite signs for the two helicities.
Since $\rygn n-(\hat{\mathbf p}\cdot\rygn n)\hat{\mathbf p}=n\mathcal K\,\hat{\boldsymbol\nu}$ by the ray equation, the Magnus term~(\ref{eq:Magnus-term}) can also be written as
\begin{equation}
\frac{\sigma}{k_0n^{2}}\,\rygn n\times\hat{\mathbf p}=-\frac{\sigma\,\mathcal K}{k_0\,n}\,\hat{\boldsymbol\beta} .
\label{eq:Magnus-binormal}
\end{equation}
The spin-induced deflection is therefore determined by the curvature of the ray alone and is directed along the binormal, perpendicular to the local plane of the ray \cite{Bliokh2009}.

\subsection{Canonical Hamiltonian}\label{subsec:Hamiltonian}

Define the canonical coordinate $\mathbf r$ by
\begin{equation}
\mathbf R=\mathbf r+\frac{1}{k_0}\,\mathbf A_\sigma(\mathbf p) .
\label{eq:R-canonical}
\end{equation}
Then $\mathbf p\cdot\mathbf R'+k_0^{-1}\mathbf A_\sigma\cdot\mathbf p'=\mathbf p\cdot\mathbf r'+k_0^{-1}\,d(\mathbf p\cdot\mathbf A_\sigma)/d\ell$, so that, up
to a total derivative, the action~(\ref{eq:phase-space-action}) is $\int d\ell\,[\mathbf p\cdot\mathbf r'-\mathcal H_\sigma]$ with
\begin{align}
\mathcal H_\sigma(\mathbf r,\mathbf p)&=|\mathbf p|-n\!\left(\mathbf r+\frac{1}{k_0}\mathbf A_\sigma(\mathbf p)\right)
\nonumber\\
&\simeq|\mathbf p|-n(\mathbf r)-\frac{1}{k_0}\,\rygn n(\mathbf r)\cdot\mathbf A_\sigma(\mathbf p) ,
\label{eq:H_tot_corrected}
\end{align}
where $(\mathbf r,\mathbf p)$ are canonical variables, $\{r_i,p_j\}=\delta_{ij}$, and the second form is the expansion to
first order in the shift $\mathbf A_\sigma/k_0$, which is of the order of a wavelength. The
Hamiltonian $\mathcal H_\sigma$ is a first-order model: terms of order $1/k_0^2$ and higher are not included, so that
it is not an exact Hamiltonian of the problem. The second term is the spin-orbit (Berry) interaction, which couples
the helicity to the gradient of the refractive index. Since $\rygn n=d\mathbf p/d\ell$ at leading order, the interaction term can also be written as $-k_0^{-1}\mathbf A_\sigma\cdot d\mathbf p/d\ell$, the
form in which the Berry phase enters the optical path length~(\ref{eq:path}).
This agrees with the Hamiltonian of Bliokh's review \cite{Bliokh2009} [Eq.~(3.11)], as discussed in Appendix~\ref{subsec:lit-conventions}.

We derive Hamilton's equations from the first form of Eq.~(\ref{eq:H_tot_corrected}). With $\partial_j=\partial/\partial R_j$,
\begin{align}
\frac{dr_i}{d\ell}&=\frac{\partial\mathcal H_\sigma}{\partial p_i}
=\hat p_i-\frac{1}{k_0}\,\partial_jn(\mathbf R)\,\frac{\partial A_{\sigma,j}}{\partial p_i} ,
\label{eq:rdot}\\
\frac{dp_i}{d\ell}&=-\frac{\partial\mathcal H_\sigma}{\partial r_i}=\partial_in(\mathbf R) ,
\label{eq:pdot}
\end{align}
where we used $\partial R_j/\partial r_i=\delta_{ij}$. The momentum feels the force at the physical position $\mathbf R$.
The term with $\partial A_{\sigma,j}/\partial p_i$ in Eq.~(\ref{eq:rdot}) is not gauge invariant. For the physical coordinate,
$dR_i/d\ell=dr_i/d\ell+k_0^{-1}(\partial A_{\sigma,i}/\partial p_j)\,dp_j/d\ell$, and
\begin{align}
\frac{dR_i}{d\ell}&=\hat p_i+\frac{1}{k_0}\,\partial_jn(\mathbf R)
\Big(\frac{\partial A_{\sigma,i}}{\partial p_j}-\frac{\partial A_{\sigma,j}}{\partial p_i}\Big)
\nonumber\\
&=\hat p_i-\frac{1}{k_0}\big[\rygn n(\mathbf R)\times\ryvOm_\sigma\big]_i ,
\label{eq:Rdot1}
\end{align}
which reproduces Eqs.~(\ref{eq:HamEqr_cor}) and (\ref{eq:HamEqp_cor}). The term of Eq.~(\ref{eq:rdot}),
$-k_0^{-1}(\rygn n\cdot\rygnp)\mathbf A_\sigma$, does not appear in the equation for $\mathbf R$ because it is
cancelled by the shift of the coordinate: this is why the equations for $d\mathbf r/d\ell$ and for $d\mathbf R/d\ell$ are different, and
why one must use the physical coordinate $\mathbf R$, and not $\mathbf r$, to describe the ray. The canonical coordinate $\mathbf r$ is not
observable\ (see Sec.~\ref{subsec:why-R}).

\subsection{Why $\mathbf R$, and not $\mathbf r$, is the position of the ray}\label{subsec:why-R}

In Secs.~\ref{sec:maxwell} and \ref{sec:berry}, $\mathbf r$ denotes a space point: the fields, the eikonal $\psi(\mathbf r)$ and the index $n(\mathbf r)$ are functions of it, and at zeroth order the ray is the curve $\mathbf r(\ell)$ of Eq.~(\ref{eq:ray0}). To this order there is only one candidate for the position of the ray. At first order in $1/k_0$, the action~(\ref{eq:phase-space-action}) and the Hamiltonian~(\ref{eq:H_tot_corrected}) involve two curves, $\mathbf R(\ell)$ and $\mathbf r(\ell)=\mathbf R(\ell)-\mathbf A_\sigma(\mathbf p)/k_0$. Both are points of the same physical space and both have the dimension of a length, but they are different curves, separated by $|\mathbf A_\sigma|/k_0\sim1/(k_0n)$, i.e., by about $\lambda/2\pi$. This is the same order as the Magnus displacement that we want to calculate, so one must decide which of the two curves describes where the light is. We choose $\mathbf R$, for the following three reasons.

\emph{(i) Gauge invariance.} The connection $\mathbf A_\sigma$ depends on the arbitrary choice of the phase of $\hat{\mathbf e}_\sigma(\mathbf p)$ [Eqs.~(\ref{eq:e1-e2}) and (\ref{eq:gauge-phase})]. Under a change of this choice, $\mathbf r\to\mathbf r+(\sigma/k_0)\rygnp\chi$, whereas $\mathbf R$ is unchanged (Sec.~\ref{subsec:gauge}). For example, the connections (\ref{eq:Berry-connection-def}) and (\ref{eq:Berry-connection-def_1}) with the same $\hat{\mathbf a}$ differ by $\sigma\rygnp\phi_{\hat a}$, so that the canonical coordinates obtained with them differ by $(\sigma/k_0)\rygnp\phi_{\hat a}$, whose magnitude $1/(k_0\,p\sin\vartheta_{\hat a})$, where $\vartheta_{\hat a}$ is the angle between $\mathbf p$ and $\hat{\mathbf a}$, is comparable to the displacement~(\ref{eq:Delta-Y}). A measured position of the light cannot depend on how the phase of the polarization basis vector is chosen. Therefore $\mathbf r$ is not an observable position, whereas $\mathbf R$ can be.

\emph{(ii) Equations of motion and conservation laws.} The velocity $d\mathbf R/d\ell$, Eq.~(\ref{eq:HamEqr_cor}), contains only the gauge-invariant curvature $\ryvOm_\sigma$, whereas $d\mathbf r/d\ell$, Eq.~(\ref{eq:rdot}), contains the gauge-dependent term $\partial A_{\sigma,j}/\partial p_i$. 
Similarly, the total angular momentum~(\ref{eq:J-conserved}), which is derived in Sec.~\ref{subsec:angmom} is gauge invariant. It has the orbital part $\mathbf R\times\mathbf p$, and substituting $\mathbf R=\mathbf r+\mathbf A_\sigma(\mathbf p)/k_0$ from Eq.~(\ref{eq:R-canonical}) gives the same conserved quantity in the canonical variables,
\begin{equation}
\mathbf J=\mathbf r\times\mathbf p+\frac{1}{k_0}\,\mathbf A_\sigma(\mathbf p)\times\mathbf p+\frac{\sigma}{k_0}\,\hat{\mathbf p} .
\label{eq:J-canonical}
\end{equation}
Some of the terms in this equation which area not gauge invariant.  Under a change of gauge, $\mathbf A_\sigma\to\mathbf A_\sigma-\sigma\nabla_{\mathbf p}\chi$, and the canonical coordinate changes as $\mathbf r\to\mathbf r+(\sigma/k_0)\nabla_{\mathbf p}\chi$ (Sec.~\ref{subsec:gauge}).  Therefore, the first term of Eq.~(\ref{eq:J-canonical}) then changes by $+(\sigma/k_0)\nabla_{\mathbf p}\chi\times\mathbf p$ and the second by $-(\sigma/k_0)\nabla_{\mathbf p}\chi\times\mathbf p$, so $\mathbf J$ is gauge invariant, whereas $\mathbf r\times\mathbf p$ alone is neither gauge invariant nor conserved \cite{J-example}.

\emph{(iii) Centroid of a wave packet.} For a wave packet of helicity $\sigma$ in a region where $n$ is approximately constant, the centroid of the intensity is $\mathbf R=\mathbf r+\mathbf A_\sigma/k_0$: the canonical coordinate $\mathbf r$ is the stationary-phase point of the scalar phase alone, and $\mathbf A_\sigma/k_0$ is the additional displacement caused by the dependence of the polarization vector on the direction of the wave vector (Appendix~\ref{app:centroid}). This is the identification used in the semiclassical wave-packet theory \cite{Sundaram1999,Onoda2004,BliokhPRE2004} and in Ref.~\cite{Bliokh2009}, where the ray is the trajectory of the centre of gravity of a paraxial beam.

\emph{Notation.} The symbol $\mathbf r$ therefore has two meanings in this paper: a point of space (the argument of the fields, of $\psi$ and of $n$), and, in Secs.~\ref{subsec:Hamiltonian}--\ref{subsec:gauge} and Appendix~\ref{sec:time}, the canonical coordinate of the ray. The two coincide only at zeroth order. The ray computed in Sec.~\ref{sec:numerical} is $\mathbf R(\ell)$, and its Cartesian components are the ray coordinates $x(\ell)$, $y(\ell)$, $z(\ell)$ shown in the figures. (Readers who prefer to reserve $\mathbf r$ for the position of the light may rename the canonical coordinate, e.g.\ $\mathbf q$; nothing else changes. Bliokh's review \cite{Bliokh2009} uses $\mathbf r$ for the physical position, which is our $\mathbf R$.)

\subsection{Check with the expanded Hamiltonian}\label{subsec:expanded-check}

If one starts from the expanded form of Eq.~(\ref{eq:H_tot_corrected}), Hamilton's equations are
\begin{align}
\frac{dr_i}{d\ell}&=\hat p_i-\frac{1}{k_0}\,\partial_jn(\mathbf r)\,\frac{\partial A_{\sigma,j}}{\partial p_i} ,\\
\frac{dp_i}{d\ell}&=\partial_in(\mathbf r)+\frac{1}{k_0}\,\partial_i\partial_jn(\mathbf r)\,A_{\sigma,j} .
\end{align}
These differ from Eqs.~(\ref{eq:rdot}) and (\ref{eq:pdot}) at order $1/k_0^{2}$, because
$\rygn n(\mathbf R)=\rygn n(\mathbf r)+k_0^{-1}(\mathbf A_\sigma\cdot\rygn)\rygn n+O(k_0^{-2})$. In particular
$d\mathbf p/d\ell=\rygn n(\mathbf R)+O(k_0^{-2})$, and Eq.~(\ref{eq:HamEqr_cor}) is obtained with an error of order $k_0^{-2}$. The expanded Hamiltonian
and the shifted-argument Hamiltonian therefore give the same ray equations to first order in $1/k_0$, which is the accuracy of the model; neither is
preferred by the approximation.

\subsection{Gauge dependence}\label{subsec:gauge}

A change of gauge,  [see Eq.~(\ref{eq:gauge-phase})], 
\begin{align}  \label{eq:gauge_transformation}
\mathbf A_\sigma\to\mathbf A_\sigma'=\mathbf A_\sigma-\sigma\rygnp\chi(\mathbf p)
\end{align}
does not change $\ryvOm_\sigma$. In the Hamiltonian~(\ref{eq:H_tot_corrected}) it changes the argument of $n$ unless the canonical coordinate is redefined,
\begin{equation}
\mathbf r'=\mathbf r+\frac{\sigma}{k_0}\rygnp\chi(\mathbf p),\qquad\mathbf p'=\mathbf p .
\end{equation}
This is a canonical transformation, generated by $F(\mathbf r,\mathbf p')=\mathbf r\cdot\mathbf p'+(\sigma/k_0)\chi(\mathbf p')$, with $\mathbf r'=\partial F/\partial\mathbf p'$ and
$\mathbf p=\partial F/\partial\mathbf r$. The Hamiltonian is form invariant, $\mathcal H_\sigma(\mathbf r,\mathbf p;\mathbf A_\sigma)=\mathcal H_\sigma(\mathbf r',\mathbf p;\mathbf A_\sigma')$,
because $\mathbf r+\mathbf A_\sigma/k_0=\mathbf r'+\mathbf A_\sigma'/k_0$. The physical position $\mathbf R$ is unchanged. In the expanded form the term
$-k_0^{-1}\rygn n\cdot\mathbf A_\sigma$ changes by $+(\sigma/k_0)\rygn n\cdot\rygnp\chi$, which is compensated, to the same order, by the shift of the canonical coordinate.
The gauge freedom therefore affects only the unobservable coordinate $\mathbf r$.

\subsection{Conservation of angular momentum}\label{subsec:angmom}

The displacement of the ray is connected to the conservation of the total angular momentum. To obtain a formula for the conserved angular momentum, we start from the orbital angular momentum of the ray, $\mathbf R\times\mathbf p$, and compute its rate of change using Eqs.~(\ref{eq:HamEqr_cor}) and (\ref{eq:HamEqp_cor}). Since $\hat{\mathbf p}\times\mathbf p=0$,
\begin{equation}
\frac{d}{d\ell}(\mathbf R\times\mathbf p)=\frac{d\mathbf R}{d\ell}\times\mathbf p+\mathbf R\times\frac{d\mathbf p}{d\ell}
=\frac{\sigma}{k_0p^{3}}\,(\rygn n\times\mathbf p)\times\mathbf p+\mathbf R\times\rygn n .
\label{eq:dLorb}
\end{equation}
For any $\mathbf b$, $(\mathbf b\times\mathbf p)\times\mathbf p=(\mathbf b \cdot \mathbf p)\,\mathbf p-p^{2}\mathbf b$, hence the first term is $-(\sigma/k_0p)\,\rygn_\perp n$, where $\rygn_\perp n=\rygn n-\hat{\mathbf p}(\hat{\mathbf p}\cdot\rygn n)$, which is the component of $\rygn n$ perpendicular to $\hat{\mathbf p}$. On the other hand, the direction of the momentum changes at the rate
\begin{equation*}
\frac{d\hat{\mathbf p}}{d\ell}=\frac{1}{p}\Big[\frac{d\mathbf p}{d\ell}-\hat{\mathbf p}\Big(\hat{\mathbf p}\cdot\frac{d\mathbf p}{d\ell}\Big)\Big]=\frac{\rygn_\perp n}{p} .
\end{equation*}
The anomalous (Magnus) velocity therefore changes the orbital angular momentum by a total derivative, $-(\sigma/k_0)\,d\hat{\mathbf p}/d\ell$, and Eq.~(\ref{eq:dLorb}) can be written as $d[\mathbf R\times\mathbf p+(\sigma/k_0)\hat{\mathbf p}]/d\ell=\mathbf R\times\rygn n$. This identifies the total angular momentum
\begin{equation}
\mathbf J=\mathbf R\times\mathbf p+\frac{\sigma}{k_0}\,\hat{\mathbf p} ,
\label{eq:J-conserved}
\end{equation}
as the sum of the orbital angular momentum and the spin angular momentum $(\sigma/k_0)\hat{\mathbf p}$, whose derivative with respect to arc-length is
\begin{equation}
\frac{d\mathbf J}{d\ell}=\mathbf R\times\rygn n(\mathbf R) .
\label{eq:dJdl}
\end{equation}
Multiplied by $\hbar k_0$, Eq.~(\ref{eq:J-conserved}) reads $\hbar k_0\mathbf J=\mathbf R\times\hbar\mathbf k+\hbar\sigma\hat{\mathbf k}$, where $\hbar\sigma\hat{\mathbf k}$ is the spin of a photon of helicity $\sigma$ \cite{classical-commutator}. This is the form used in Sec.~\ref{sec:interface}.

Hence $\mathbf J$ is exactly conserved in a spherically symmetric medium, $n=n(|\mathbf R|)$.  Moreover, for a medium that is symmetric under rotations about an axis, the component of $\mathbf J$ along this axis is conserved. This is the integral of motion of Ref.~\cite{Bliokh2009} [Eq.~(3.42)], $\mathbf J=\mathbf r\times\mathbf p+\bar\lambda_0s\,\mathbf t$, written with the physical position $\mathbf R$. The unit tangent $\mathbf t=d\mathbf R/d\ell$ differs from $\hat{\mathbf p}$ only at order $1/k_0$, so the two spin terms agree to the order of the theory. The interface example of Sec.~\ref{sec:interface} is the case of the axis along the normal to the interface.

\section{Size of the Magnus effect and validity}\label{sec:validity}

Integrating the Magnus term of Eq.~(\ref{eq:HamEqr_cor}) along a ray and using $\rygn n\,d\ell=d\mathbf p$, the part of the transverse displacement that comes directly from the Magnus (anomalous-velocity) term is
\begin{equation}
\Delta\mathbf R_{\sigma,\rm M}=-\frac{1}{k_0}\int d\mathbf p\times\ryvOm_\sigma(\mathbf p)
=\frac{\sigma}{k_0}\int d\mathbf p\times\frac{\mathbf p}{p^{3}} ,
\label{eq:Delta-R}
\end{equation}
where the integral is along the path of $\mathbf p$ in momentum space. This contribution does not depend on the details of $n(\mathbf r)$, whereas the
zeroth-order displacement $\int\hat{\mathbf p}\,d\ell$ does. If the direction of $\mathbf p$ turns through an angle $\Delta\phi_\sigma$ (the turning angle of the ray of helicity $\sigma$) in a plane while $n$ changes
little, the displacement is perpendicular to this plane and has the magnitude $|\Delta R_{\sigma,\rm M}|\simeq\Delta\phi_\sigma/(k_0n)=\lambda\,\Delta\phi_\sigma/2\pi$, with
$\lambda=\lambda_0/n$ the wavelength in the medium. The turning angle depends on $\sigma$ only through the Magnus displacement of the ray, which changes the force $\rygn n(\mathbf R)$ that acts on $\mathbf p$ [Eq.~(\ref{eq:HamEqp_cor})]:
\begin{equation}
\Delta\phi_\sigma=\Delta\phi_0+O(1/k_0),
\label{eq:Dphi-sigma}
\end{equation}
where $\Delta\phi_0$ is the turning angle of the reference ray ($\sigma=0$, Sec.~\ref{subsec:num-equations}). The correction is small, of relative order $1/(k_0L)$, and, because $\Delta R_{\sigma,\rm M}$ is itself of order $1/k_0$, it changes $\Delta R_{\sigma,\rm M}$ only at order $1/k_0^2$, beyond the accuracy of this first-order estimate; to this order $\Delta\phi_\sigma$ may be replaced by $\Delta\phi_0$. It is of the order of a wavelength times the turning angle.  

Equation~(\ref{eq:Delta-R}) is, however, only the direct contribution of the anomalous-velocity term. To first order in $1/k_0$, the full displacement $\delta\mathbf R$ of the ray with helicity $\sigma$ from the ray with $\sigma=0$ that starts with the same initial conditions obeys, from Eqs.~(\ref{eq:HamEqr_cor}) and (\ref{eq:HamEqp_cor}) linearized about the $\sigma=0$ ray,
\begin{align}
\frac{d\,\delta\mathbf R}{d\ell}&=\frac{\delta\mathbf p_\perp}{n}+\frac{\sigma}{k_0n^{2}}\,\rygn n\times\hat{\mathbf p},
\label{eq:first-order-response}\\
\frac{d\,\delta\mathbf p}{d\ell}&=(\delta\mathbf R\cdot\rygn)\rygn n ,\nonumber
\end{align}
where $\delta\mathbf p_\perp$ is the component of $\delta\mathbf p$ perpendicular to $\hat{\mathbf p}$ (this is Eq.~(3.44) of Ref.~\cite{Bliokh2009}). The first term of Eq.~(\ref{eq:first-order-response}) is the response of the ray to its displacement in the inhomogeneous medium. Equation~(\ref{eq:Delta-R}) is obtained by dropping it, i.e., by setting $\delta\mathbf p=0$ [Eq.~(3.45) of the review \cite{Bliokh2009}]. The response vanishes if $n$ is invariant under translations along $\delta\mathbf R$, as for the planar interface of Sec.~\ref{sec:interface}, but not in general: it changes the displacement of a helical ray by a factor $1/2$ (Appendix~\ref{subsec:lit-analytic}), and it contributes substantially to the out-of-plane displacement of the rays of Sec.~\ref{subsec:num-results}.

The following remarks concern the validity of the description.
\begin{enumerate}
\item The ray equations are derived to first order in $1/(k_0L)$, where $L$ is the length scale of the variation of $n$ and $\epsilon$, $\mu$. The
Hamiltonian~(\ref{eq:H_tot_corrected}) is a first-order model and it is not an exact Hamiltonian of the problem. Terms of second order, such as the quadratic term in the expansion of
$n(\mathbf r+\mathbf A_\sigma/k_0)$, are of the same order as other neglected corrections, and have no separate physical meaning.
\item The shift $\mathbf A_\sigma/k_0$ is not observable by itself because it depends on the gauge; only $\mathbf R$ and the curvature are meaningful.
The identification of $\mathbf R$ with the center of the wave packet is taken from the semiclassical theory \cite{Sundaram1999,Onoda2004,BliokhPRE2004}; a short derivation for a locally homogeneous region is given in Appendix~\ref{app:centroid}.
\item The sign of the Magnus term depends on the convention for the helicity $\sigma$ and for the sign of the curvature; Eq.~(\ref{eq:Berry-curvature}) with $\sigma=+1$
for positive helicity fixes it here.
\item The polarization analysis neglects the transverse derivatives of the amplitudes (see Sec.~\ref{subsec:first-order}) and
assumes $\epsilon,\mu$ real. Coupling between the two helicity states is not present to the order considered (Sec.~\ref{subsec:circular}); it is of higher order in
$1/(k_0L)$ for a smoothly varying medium.
\item For linear or elliptical polarization the description requires the superposition of the two helicity rays (spin Hall splitting).
\end{enumerate}

\section{Example: refraction at a planar interface}\label{sec:interface}

As a check, consider a refractive index that depends on $z$ only,
\begin{equation}
n(z)=
\begin{cases}
n_1, & z<-a ,\\
\tilde n(z), & -a\le z\le a ,\\
n_2, & z>a ,
\end{cases}
\end{equation}
where $a>0$ is the half-width of the interface and $\tilde n(z)$ is continuous with $\tilde n(-a)=n_1$ and $\tilde n(a)=n_2$; for example
$\tilde n(z)=\tfrac12(n_1+n_2)-\tfrac12(n_1-n_2)z/a$. A ray in the $x$-$z$ plane (the plane of incidence) arrives from $z<-a$ with the angle
of incidence $\theta_i$. The component $p_x=n_1\sin\theta_i$ is conserved (Snell's law), and
\begin{equation}
\mathbf p=\Big(s,\,0,\,\sqrt{n^2-s^2}\Big),\, \, \, \text{where  } s \equiv n_1 \sin\theta_i = n\sin\theta .
\label{eq:p-vs-n}
\end{equation}
Here $\theta$ is the local angle between the ray and the $z$ axis.  Since $\rygn n=n'(z)\,\hat{\mathbf z}$ and
$\hat{\mathbf z}\times\hat{\mathbf p}=\sin\theta\,\hat{\mathbf y}$, the Magnus term of Eq.~(\ref{eq:HamEqr_cor}) is along $\hat{\mathbf y}$, perpendicular to the plane of incidence,
\begin{equation}
\frac{dY}{d\ell}=\frac{\sigma\,n'(z)\sin\theta}{k_0n^{2}},
\end{equation}
where $Y$ is the $y$ component of $\mathbf R$. Since the Magnus term has no $z$ component here, Eq.~(\ref{eq:HamEqr_cor}) gives $dz/d\ell=\cos\theta$ exactly, so that $dY=\sigma\tan\theta\,dn/(k_0n^2)$ with $\tan\theta=s/\sqrt{n^2-s^2}$, and the integral gives
\begin{equation}
\Delta Y=\frac{\sigma}{k_0}\int_{n_1}^{n_2}\frac{s\,dn}{n^{2}\sqrt{n^{2}-s^{2}}}
=\frac{\sigma\,(\cos\theta_t-\cos\theta_i)}{k_0\,n_1\sin\theta_i} ,
\label{eq:Delta-Y}
\end{equation}
where we have used $\int \frac{dn}{n^2\sqrt{n^2 - s^2}} = \frac{\sqrt{n^2 - s^2}}{n s^2}$, $s = n_1 \sin\theta_i$, and $\sin\theta_t = (n_1/n_2)\sin\theta_i$. The shift is independent of the profile $\tilde n(z)$ and of $a$, vanishes at normal incidence, and is of the order of a wavelength.  Because $\mathbf p$ stays in the $x$-$z$ plane and $n$ does not depend on $y$, the displacement along $y$ produces no response of the momentum [the first term of Eq.~(\ref{eq:first-order-response}) vanishes], so that Eq.~(\ref{eq:Delta-Y}) is exact within Eqs.~(\ref{eq:HamEqr_cor}) and (\ref{eq:HamEqp_cor}).  For example, for $n_1=1$, $n_2=1.5$ and $\theta_i=60^\circ$, $\Delta Y=0.366\,\sigma/k_0=0.058\,\sigma\lambda_0$.

Equation~(\ref{eq:Delta-Y}) coincides with what follows from the conservation of the component of the total angular momentum along the normal to the
interface, $J_z=(\mathbf R\times\hbar\mathbf k)_z+\hbar\sigma\cos\theta$, where the second term is the spin angular momentum of a photon of helicity $\sigma$ (spin along $\hat{\mathbf p}$).
With $k_y=0$ and $k_x=k_0n_1\sin\theta_i$ conserved, $J_z/\hbar=-Yk_x+\sigma\cos\theta$ gives $\Delta Y=\sigma\,\Delta\cos\theta/k_x$, which is Eq.~(\ref{eq:Delta-Y}). This confirms
the sign convention of Eq.~(\ref{eq:Berry-curvature}) with $\sigma=+1$ for positive helicity.

The Berry phase is gauge dependent and is not an observable by itself. In the gauge~(\ref{eq:Berry-connection-def_1}) with $\hat{\mathbf a}=\hat{\mathbf y}$, which is perpendicular to the plane of
incidence, $\tilde{\mathbf A}_\sigma=\sigma(-p_z,0,p_x)/p^2$ and $d\mathbf p=(0,0,dp_z)$, so that
\begin{align}
\int\tilde{\mathbf A}_\sigma\cdot d\mathbf p&=\sigma\int_{n_1}^{n_2}\frac{s\,dn}{n\sqrt{n^2-s^2}}\nonumber\\
&=\sigma\,[\arccos(s/n)]_{n_1}^{n_2}=\sigma\,(\theta_i-\theta_t) .
\end{align}
This is the gauge term $-\sigma\Delta\phi_{\hat a}$ of Eq.~(\ref{eq:gauge-phase}): $\Delta\phi_{\hat a}=\theta_t-\theta_i$ is the change of the azimuth of $\mathbf p$ around
$\hat{\mathbf a}$. In the gauge~(\ref{eq:Berry-connection-def}) with the same $\hat{\mathbf a}$, $\mathbf p\cdot\hat{\mathbf a}=0$ and $\mathbf A_\sigma=0$ on this path, and the torsion of the planar ray vanishes, so that the Frenet phase~(\ref{eq:Berry-phase-frenet}) is also zero.
The observable effect is the displacement~(\ref{eq:Delta-Y}), which depends only on the curvature.

The spin Hall effect in a gradient-index medium and the transverse Imbert-Fedorov shift of a beam reflected or transmitted at a plane interface are the weak- and strong-scattering limits of the same polarization transport \cite{Bliokh2009}. For a low contrast the Imbert-Fedorov shift corresponds to the ray equations~(\ref{eq:HamEqr_cor}); for a general contrast the evolution at a sharp interface is not adiabatic and the shift cannot be described by the Berry-phase formalism \cite{BliokhIF2006,HostenKwiat2008}. Equation~(\ref{eq:Delta-Y}) applies to a graded interface whose width is large compared with the wavelength. For a small change $\delta\theta=\theta_t-\theta_i$ of the angle it gives $\Delta Y\simeq-\sigma\,\delta\theta/(k_0n_1)$, a wavelength times the angle of deflection divided by $2\pi$, as expected from the estimate of
Sec.~\ref{sec:validity}.

\section{Numerical propagation of rays}\label{sec:numerical}

\subsection{Equations to be integrated}\label{subsec:num-equations}

The ray of helicity $\sigma$ is obtained by integrating Eqs.~(\ref{eq:HamEqr_cor}) and (\ref{eq:HamEqp_cor}) in the variables $(\mathbf R,\mathbf p)$,
\begin{align}
\frac{d\mathbf R}{d\ell}&=\frac{\mathbf p}{|\mathbf p|}+\frac{\sigma}{k_0}\,\frac{\rygn n(\mathbf R)\times\mathbf p}{|\mathbf p|^{3}} ,
\label{eq:num-R}\\
\frac{d\mathbf p}{d\ell}&=\rygn n(\mathbf R) ,
\label{eq:num-p}
\end{align}
with the initial conditions $\mathbf R(0)=\mathbf R_0$ and $\mathbf p(0)=n(\mathbf R_0)\,\hat{\mathbf p}_0$, which places the ray on the surface $|\mathbf p|=n(\mathbf R)$.
No gauge choice is needed, because only the curvature enters. Along the exact solution of Eqs.~(\ref{eq:num-R}) and (\ref{eq:num-p}), $|\mathbf p|-n(\mathbf R)$ is
conserved, which provides a test of the numerical integration. The ray coordinates obtained in this way are the components of the physical position, $\mathbf R(\ell)=(X(\ell),Y(\ell),Z(\ell))$; we do not compute the canonical coordinate $\mathbf r$, which would require a choice of gauge (Sec.~\ref{subsec:why-R}). Setting $\sigma=0$ in Eqs.~(\ref{eq:num-R}) and (\ref{eq:num-p}) switches the Magnus term off and gives the \emph{reference ray}: the ray of ordinary geometrical optics, Eq.~(\ref{eq:ray0}). No light has $\sigma=0$, but to this order the reference ray is the trajectory of the centroid of a linearly polarized beam (for which $S_3=0$ in Eq.~(3.41) of Ref.~\cite{Bliokh2009}), whose two circular components, $\sigma=\pm1$, are displaced symmetrically about it. The difference of the rays with $\sigma=+1$ and $\sigma=-1$ is the spin Hall splitting.

\subsection{Medium with a localized refractive-index profile}\label{subsec:num-medium}

As an example, we consider the index
\begin{equation}
n(x,y,z)=n_0+\Delta n\,\exp\Big[-\frac{x^2+z^2}{2w_1^2}-\frac{y^2}{2w_2^2}\Big] ,
\label{eq:n_xyz}
\end{equation}
with $n_0=1$, $\Delta n=2$, $w_1=8$ and $w_2=4$ (in units of length; $k_0$ is in the inverse units). Its gradient is
\begin{equation}
\rygn n=-\big(n-n_0\big)\Big(\frac{x}{w_1^2},\ \frac{y}{w_2^2},\ \frac{z}{w_1^2}\Big) .
\end{equation}

\subsection{Tests of the numerical calculation}\label{subsec:num-tests}

The following properties of Eqs.~(\ref{eq:num-R}) and (\ref{eq:num-p}) can be used to test an integration; the first two are exact, and the others reproduce the analytic results of Sec.~\ref{sec:interface} and Appendix~\ref{subsec:lit-analytic}.
\begin{enumerate}
\item \emph{Conservation of $|\mathbf p|-n(\mathbf R)$.} This quantity is exactly conserved (Sec.~\ref{subsec:phase-space}), for any profile.
\item \emph{Conservation of angular momentum.} The profile~(\ref{eq:n_xyz}) is invariant under rotations about the $y$ axis, so that, by Eq.~(\ref{eq:J-conserved}), the component
\begin{equation}
J_y=Zp_x-Xp_z+\frac{\sigma}{k_0}\,\frac{p_y}{|\mathbf p|}
\label{eq:Jy}
\end{equation}
is exactly conserved. For a spherically symmetric profile the full vector~(\ref{eq:J-conserved}) is conserved. The spin term is of order $1/k_0$, so that a test of Eq.~(\ref{eq:Jy}) with the spin term omitted fails at this order.
\item \emph{Helical ray.} For a cylindrically symmetric profile $n(\rho)$ [not the profile~(\ref{eq:n_xyz})], for example $n(\rho)=n_a(1-\rho^2/2\Lambda^2)$, where $n_a$ is the index on the axis and $\Lambda$ a length, for which $\sin^2\theta=(\rho^2/\Lambda^2)/(1-\rho^2/2\Lambda^2)$, let $\mathbf R_0=(\rho,0,0)$. For $\sigma=0$, $\mathbf p_0=n(\rho)\,(0,\sin\theta,\cos\theta)$ gives a helix of radius $\rho$ about the $Z$ axis. For $\sigma=\pm1$ the exact helix of radius $\rho$ has $\mathbf p_0=n(\rho)\,(0,\sin\theta_p,\cos\theta_p)$ with $\theta_p\simeq\theta-\sigma\mathcal K/(2k_0n)$ (Appendix~\ref{subsec:lit-analytic}); the ray started with this momentum is displaced from the $\sigma=0$ helix, after one period $\ell=2\pi\rho/\sin\theta$, by $\delta\mathbf R_0$ of Eq.~(\ref{eq:helix-shift}), where $\hat{\boldsymbol\beta}=(0,-\cos\theta,\sin\theta)$ at the starting point (and rotates with the helix). The displacement grows linearly with the number of periods, and the splitting between $\sigma=+1$ and $\sigma=-1$ is $\lambda\sin\theta$ per period. If both rays are started with the same momentum $\mathbf p_0$, an oscillation of the radius of the same order in $1/k_0$ (an epicycle) is superposed on this drift.
\item \emph{Graded interface.} For $n=n(z)$ as in Sec.~\ref{sec:interface}, the transverse displacement $\Delta Y$ of Eq.~(\ref{eq:Delta-Y}) is independent of the profile of the interface, and tests the integration at oblique incidence.
\end{enumerate}
Test 3 is a first-order result, reproduced by the numerical solution up to terms of order $k_0^{-2}$, so that the comparison should be made for $k_0\rho$ large and with the difference of the rays with $\sigma=\pm1$; test 4 is exact within the equations (see Sec.~\ref{sec:interface}).

\subsection{Results for the Gaussian profile}\label{subsec:num-results}

\begin{figure}
\includegraphics[width =0.97 \linewidth]{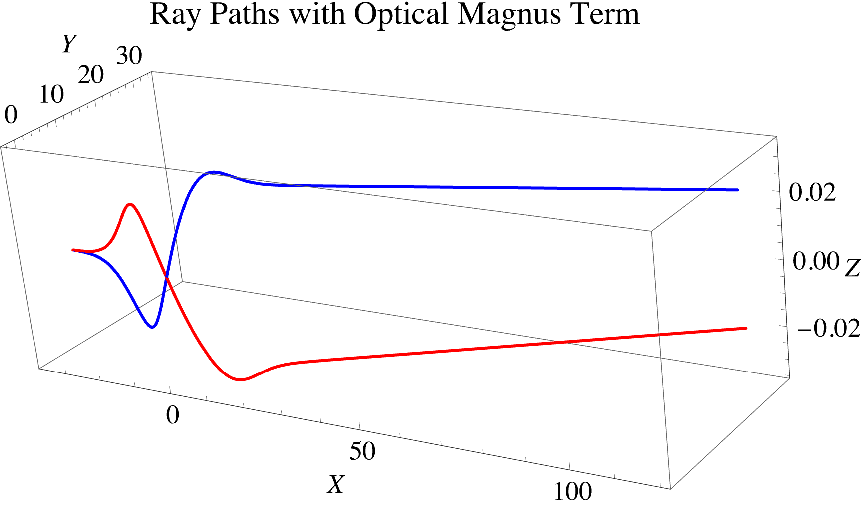}
\caption{\label{Magnus_ray_paths}
Parametric plot of the ray path ${\mathbf R}(\ell)$ for circular polarization $\sigma = +1$ (blue) and $\sigma = -1$ (red).  Note the small scale of the $z$-axis.}
\end{figure}

\begin{figure}
\includegraphics[width =0.97 \linewidth]{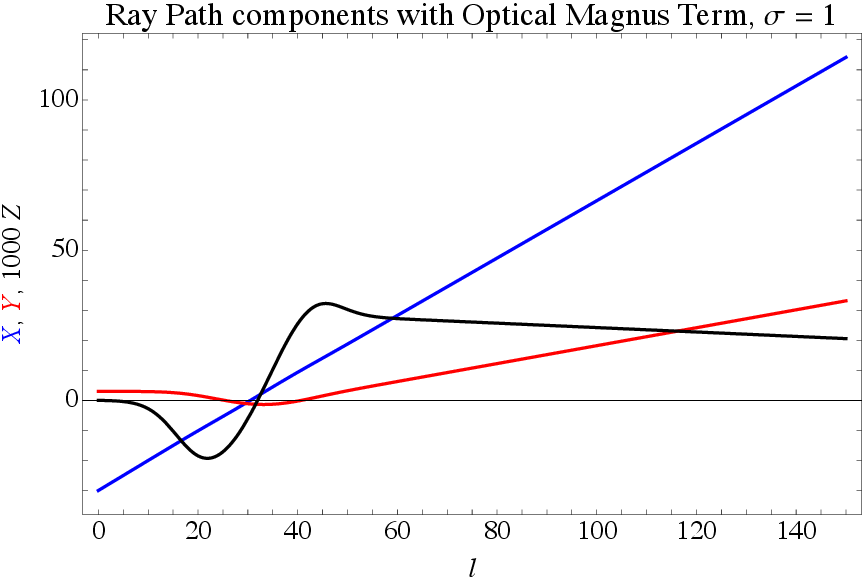}
\caption{\label{Fig_r_vs_ell_simga+1}
The ray path $X(\ell)$ (blue), $Y(\ell)$ (red), and $Z(\ell)$ (black) for circular polarization $\sigma = +1$. Note that $Z(\ell)$ is multiplied by 1000.  For $\sigma = -1$, $X$ and $Y$ are identical and $Z$ is of opposite sign.}
\end{figure}

\begin{figure}
\includegraphics[width =0.97 \linewidth]{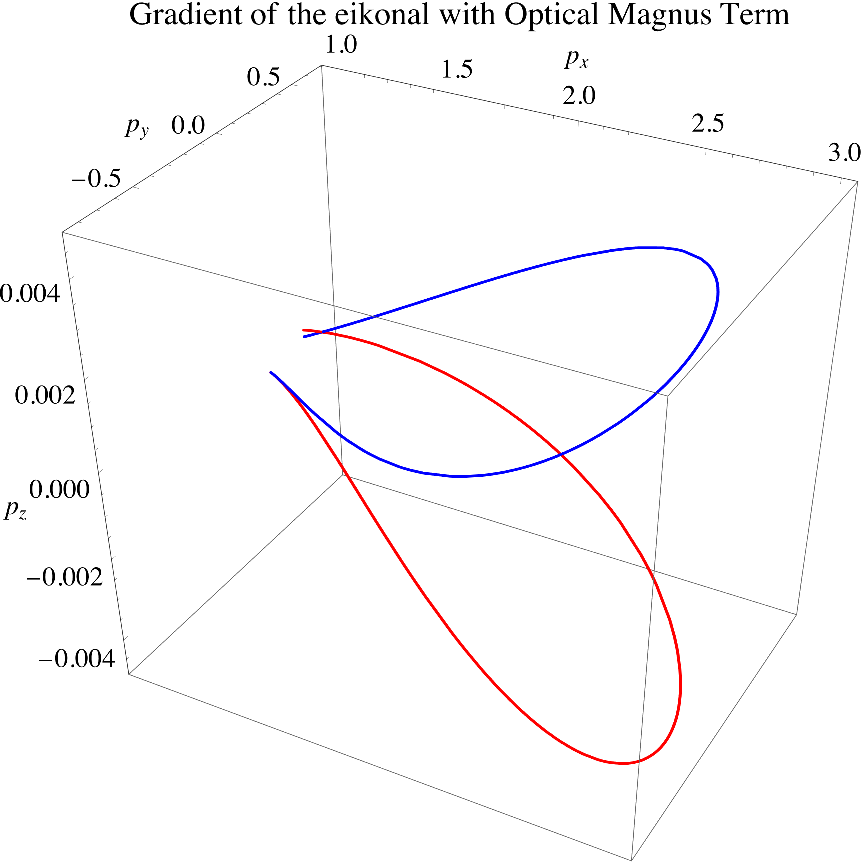}
\caption{\label{Magnus_eikonal_gradiant}
Parametric plot of the momentum, i.e. the gradient of the eikonal ${\mathbf p}(\ell) = \rygn\psi(\ell)$ for circular polarization $\sigma = +1$ (blue) and $\sigma = -1$ (red).  Note the small scale of the $p_z$ axis.}
\end{figure}

\begin{figure}
\includegraphics[width =0.97 \linewidth]{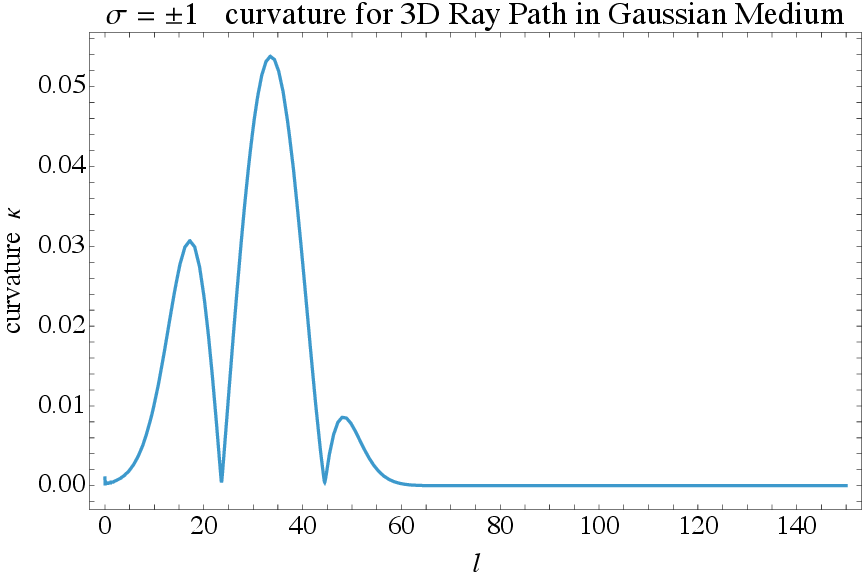}
\caption{\label{Magnus-curvature}
The ray curvature $\mathcal K=|d\hat{\mathbf p}/d\ell|$ versus path length $\ell$.  The curvature for right and left circular polarizations are equal.}
\end{figure}

\begin{figure}
\includegraphics[width =0.97 \linewidth]{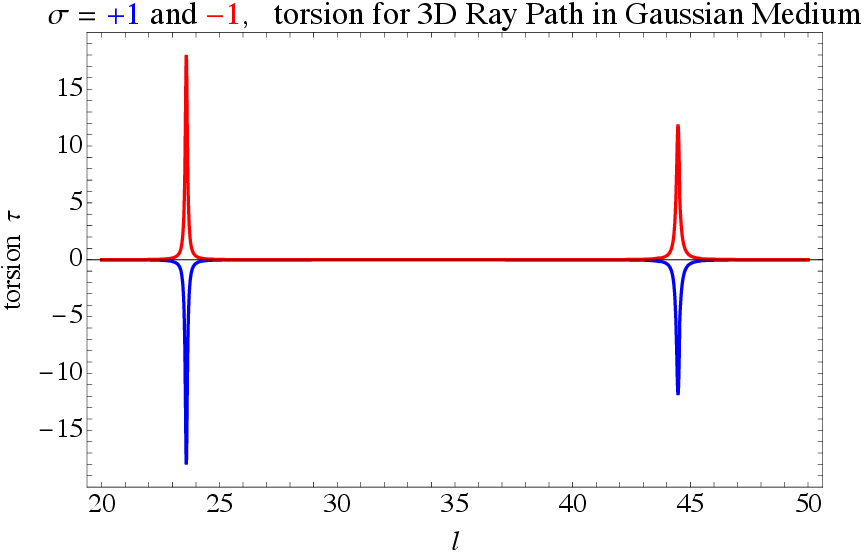}
\caption{\label{Magnus-torsion}
The torsion of the ray $\mathcal T=-\hat{\boldsymbol\nu}\cdot d\hat{\boldsymbol\beta}/d\ell$ versus path length $\ell$.  The torsion for right and left circular polarizations are opposite in sign.}
\end{figure}

We now present the numerical results of the integration of the equations of motion, Eqs.~(\ref{eq:num-R}) and (\ref{eq:num-p}), for the profile~(\ref{eq:n_xyz}): rays for $\sigma=\pm1$, with initial conditions $X(0) = -30, Y(0) = 3, Z(0) = 0$, $p_x(0)=n(-30,3,0)$, $p_y(0) = 0$, 
$p_z(0)= 0$, and $k_0 = 10$. The ray paths, their components, the momentum, and the curvature and torsion of the rays are shown in Figs.~\ref{Magnus_ray_paths}--\ref{Magnus-torsion}.

Figures~\ref{Magnus_ray_paths} and \ref{Fig_r_vs_ell_simga+1} show the ray path ${\mathbf R}(\ell)$ for circular polarization when the spatial dependence of the refractive index is given by Eq.~(\ref{eq:n_xyz}). Figure~\ref{Magnus_ray_paths} plots the rays versus $\ell$ for $\sigma = \pm 1$, and Fig.~\ref{Fig_r_vs_ell_simga+1} shows the Cartesian components of the ray path, ${\mathbf R}(\ell) = (X(\ell), Y(\ell), Z(\ell))$, versus $\ell$ for $\sigma = +1$.  Note that $Z(\ell)$ has been multiplied by 1000 --- there is only the small deviation of the rays from the $x$-$y$ plane.  This small deviation of the rays from the $x$-$y$ plane is due to the optical Magnus effect. Neglecting the Magnus effect, the ray path would lie in the $x$-$y$ plane.

Figure~\ref{Magnus_eikonal_gradiant} shows the momentum ${\mathbf p}(\ell)$ for $\sigma = \pm 1$.  The $x$- and $y$-components of ${\mathbf p}(\ell)$ are the same for the left- and right-polarized light, whereas the $z$-components are small and have opposite signs for $\sigma = \pm 1$.  The nonzero value of $p_z(\ell)$ manifests the Magnus effect.

We checked that for $k_0=20$, instead of $k_0=10$, $Z(\ell)$ and $p_z(\ell)$ are smaller by a factor of $1/2$, and more generally, increasing $k_0$ by a factor $\zeta$ makes $Z(\ell)$ and $p_z(\ell)$ smaller by a factor of $1/\zeta$.

Figure~\ref{Magnus-curvature} shows the ray curvature ${\mathcal K}(\ell)$ as a function of $\ell$, and Fig.~\ref{Magnus-torsion} shows the ray torsion ${\mathcal T}_\sigma(\ell)$. The ray curvature in Fig.~\ref{Magnus-curvature} is the same for the left- and right-polarized rays.  It has three peaks, at $\ell = 16.9$, $33.1$ and $48$, as well as two zeros at $\ell = 23.2$ and $44.0$. The ray torsion in Fig.~\ref{Magnus-torsion}  satisfies the equality ${\mathcal T}_{-1}(\ell) = -{\mathcal T}_1(\ell)$. ${\mathcal T}_1(\ell)$ has two sharp peaks at $\ell = 23.2$ and $44.0$ (the curvature vanishes at these points), and it is very small far from these points.

The out-of-plane displacement has a simple origin. For $\sigma=0$ (the reference ray of Sec.~\ref{subsec:num-equations}) the ray stays in the plane $z=0$, because $\mathbf p$ and $\rygn n$ lie in that plane. The Magnus term, proportional to $\rygn n\times\mathbf p$, is then along $\hat{\mathbf z}$, and the reflection $z\to-z$ combined with $\sigma\to-\sigma$ maps the equations into themselves, which is why $X$, $Y$, $p_x$ and $p_y$ are the same for $\sigma=\pm1$, while $Z$ and $p_z$ change sign. To first order in $1/k_0$, Eq.~(\ref{eq:first-order-response}) reduces, along the $\sigma=0$ ray, to
\begin{equation}
\frac{d\,\delta Z}{d\ell}=\frac{\delta p_z}{n}-\frac{\sigma}{k_0n}\,\frac{d\varphi_p}{d\ell},\qquad
\frac{d\,\delta p_z}{d\ell}=-\frac{n-n_0}{w_1^2}\,\delta Z ,
\label{eq:Z-linear}
\end{equation}
where $\varphi_p$ is the angle of $(p_x,p_y)$. The second equation is the focusing of the ray toward the plane $z=0$, where $n$ is maximal. For $k_0=10$, the anomalous-velocity term alone [Eq.~(\ref{eq:Delta-R})] gives $\max|Z|\simeq0.023$ and $Z(\ell=140)\simeq+0.004$, whereas the solution of the full equations has $\max|Z|\simeq0.032$ and $Z(140)\simeq-0.013$; Eq.~(\ref{eq:Z-linear}) reproduces the full solution to better than $10^{-4}$ in relative terms. The response to the displacement is thus essential for the out-of-plane displacement of the ray.

The torsion peaks of Fig.~\ref{Magnus-torsion} are located at the zeros of the curvature, i.e., at the inflection points of the ray. The $\sigma=0$ ray is planar and has $\mathcal T=0$ identically; the torsion of the rays with $\sigma=\pm1$ is nonzero only because of the small out-of-plane displacement. At a zero of the curvature the Frenet frame is not defined, $\hat{\boldsymbol\nu}$ and $\hat{\boldsymbol\beta}$ change sign, and the torsion has a spike whose integral is $\pm\pi$ independent of $k_0$ (numerically $\int\mathcal T\,d\ell\simeq3.06$ and $3.03$ for the two spikes at $k_0=10$). The Berry phase $\sigma\int\mathcal T\,d\ell$ in the Frenet gauge is therefore a multiple of $\pi$ and is a gauge artifact: for a planar ray the Berry phase vanishes in a smooth gauge, and the transport law~(\ref{eq:parallel-transport}) itself is independent of the frame.

\begin{figure}
\includegraphics[width =0.97 \linewidth]{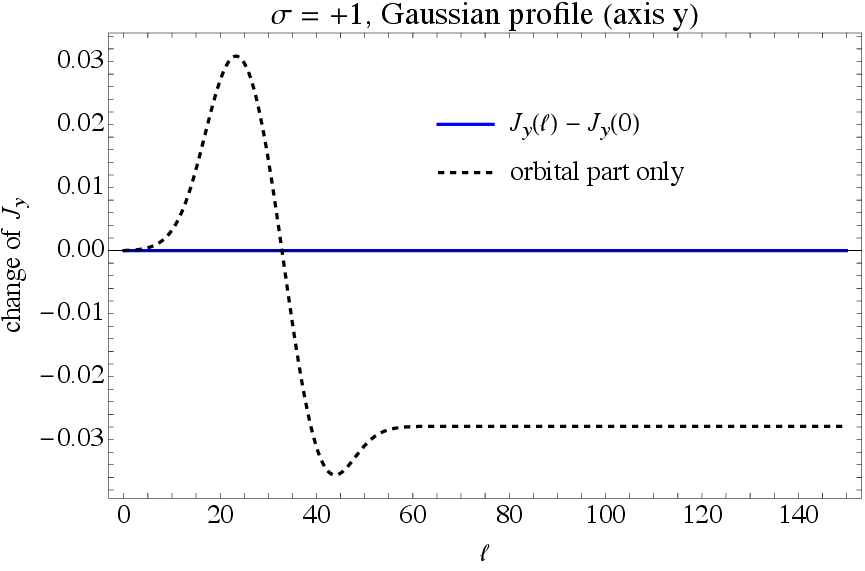}
\caption{\label{Magnus_change_Jy}
Change $\Delta J_y(\ell)=J_y(\ell)-J_y(0)$ of the component of the total angular momentum of Eq.~(\ref{eq:Jy}) (solid line, zero within the numerical accuracy) and of its orbital part $Zp_x-Xp_z$ alone (dashed line), for $\sigma=+1$.}
\end{figure}

Figure~\ref{Magnus_change_Jy} shows the change of the component $J_y$ of Eq.~(\ref{eq:Jy}) along the ray (solid line), which vanishes to within the numerical accuracy ($\sim10^{-13}$ in our check of the integration), and the change of its orbital part $Zp_x-Xp_z$ alone (dashed line), which varies by about $\pm0.03$; the difference is the spin term $(\sigma/k_0)\,p_y/|\mathbf p|$. The quantity $|\mathbf p|-n(\mathbf R)$ is conserved at the level of $10^{-12}$.

\section{Summary and conclusions}\label{sec:summary}

Starting from Maxwell's equations in an inhomogeneous medium and the eikonal expansion to first order in $1/k_0$, we showed that the first-order fields exist only if the polarization
vector is parallel transported along the ray (Sec.~\ref{subsec:first-order}). For circular polarization the two helicity states are uncoupled and acquire the Berry phase
$\varphi_\sigma=\int\mathbf A_\sigma\cdot d\mathbf p$, where the Berry connection $\mathbf A_\sigma$ has the gauge-invariant curvature $\ryvOm_\sigma=-\sigma\mathbf p/p^3$
(Sec.~\ref{sec:berry}). Including the Berry phase in the optical path length gives the action~(\ref{eq:phase-space-action}), a canonical Hamiltonian
$\mathcal H_\sigma=|\mathbf p|-n(\mathbf r+\mathbf A_\sigma/k_0)\simeq|\mathbf p|-n(\mathbf r)-k_0^{-1}\rygn n\cdot\mathbf A_\sigma$, and a Lagrangian that depends on the acceleration of the ray
(Sec.~\ref{sec:action} and Appendix~\ref{subsec:Lagrangian}). In terms of the physical position $\mathbf R=\mathbf r+\mathbf A_\sigma/k_0$ the equations of motion are
\begin{align}
\frac{d\mathbf R}{d\ell}&=\hat{\mathbf p}+\frac{\sigma}{k_0}\,\rygn n(\mathbf R)\times\frac{\mathbf p}{p^3},&
\frac{d\mathbf p}{d\ell}&=\rygn n(\mathbf R),
\end{align}
which contain only the gauge-invariant curvature. The second term of the first equation is the optical Magnus term, transverse to $\rygn n$ and to the ray, of the order of a wavelength
times the turning angle of the ray, and of opposite sign for the two helicities. The equations are equivalent, to first order in $1/k_0$, to the shifted-argument and the expanded Hamiltonians,
to the non-canonical bracket form, and to the formulation with the Hamiltonian $\hbar\omega=\hbar ck/n$ (Appendix~\ref{sec:time}). For refraction at a planar interface the Magnus term gives
the displacement~(\ref{eq:Delta-Y}), which agrees with the conservation of angular momentum about the normal to the interface (Sec.~\ref{sec:interface}). The equations are in a form suitable for numerical propagation
(Sec.~\ref{sec:numerical}). The Magnus term is the helicity-dependent transverse deflection; it scales as $1/k_0$ and vanishes in the limit $k_0\to\infty$, leaving the conventional bending of the ray toward optically denser regions.

The Berry phase and the spin Hall effect are consequences of the same spin-orbit term, but they differ in their dependence on the frame: the Berry phase of an open path depends on the gauge (Sec.~\ref{subsec:Berry-curvature}, Appendix~\ref{app:frenet} and the interface example), whereas the displacement of the ray is gauge invariant and is determined by the curvature of the ray, Eq.~(\ref{eq:Magnus-binormal}). The two helicity states are uncoupled to first order; an elliptically polarized beam splits upon propagation into two overlapping but displaced circularly polarized beams \cite{Bliokh2009}, each following Eqs.~(\ref{eq:HamEqr_cor}) and (\ref{eq:HamEqp_cor}). The Berry-phase description requires a smooth medium; sharp interfaces give the non-adiabatic Imbert-Fedorov shift, which is not contained in the equations of this paper.

The equations agree with those of Bliokh's review \cite{Bliokh2009}, including the sign of the Magnus term (Appendix~\ref{subsec:lit-conventions}). They reproduce the known analytic results: the deflection along the binormal determined by the curvature of the ray, Eq.~(\ref{eq:Magnus-binormal}); the shift of a helical ray, Eq.~(\ref{eq:helix-shift}); the conservation of the total angular momentum, Eq.~(\ref{eq:J-conserved}); and the displacement at a graded interface, Eq.~(\ref{eq:Delta-Y}), which agrees with the low-contrast limit of the Imbert-Fedorov shift (Sec.~\ref{sec:interface}). Experiments confirm the ray equations for helical rays \cite{Bliokh2008,Bliokh2015}, and the same equations have been integrated numerically for light in the Schwarzschild spacetime \cite{Oancea2020} and for vortex light in a waveguide \cite{Qiu2024}. We did not find published numerical propagation for a general three-dimensional gradient-index profile; the exact invariants and the analytic tests of Sec.~\ref{subsec:num-tests} are provided for such a calculation.

\appendix

\section{First-order solvability condition}\label{app:transport}

Here we evaluate the vector $\boldsymbol{\mathcal C}$ of Eq.~(\ref{eq:B-vector}) for the zeroth-order fields~(\ref{eq:e_0-general}) and (\ref{eq:h_0-general}), using Eq.~(\ref{eq:curl-ray}) and the Frenet-Serret formulas~(\ref{eq:Frenet-Serret-formulas}).
Using Eqs.~(\ref{eq:Frenet-Serret-formulas}) one finds
\begin{equation}
\frac{d\mathbf E_0}{d\ell}=D_\nu\,\hat{\boldsymbol\nu}+D_\beta\,\hat{\boldsymbol\beta}-\mathcal K E_\nu\,\hat{\mathbf p} ,
\label{eq:dE0}
\end{equation}
with
\begin{equation}
D_\nu=\frac{dE_\nu}{d\ell}-\mathcal T E_\beta ,
\qquad
D_\beta=\frac{dE_\beta}{d\ell}+\mathcal T E_\nu .
\label{eq:D-def}
\end{equation}
The curls follow from Eq.~(\ref{eq:curl-ray}),
\begin{align}
\rygn\times\mathbf E_0&=-D_\beta\,\hat{\boldsymbol\nu}+D_\nu\,\hat{\boldsymbol\beta},
\label{eq:curl-e0}\\
c\mu\,\rygn\times\mathbf H_0&=-n\,D_\nu\,\hat{\boldsymbol\nu}-n\,D_\beta\,\hat{\boldsymbol\beta}
-n\,\frac{d\ln\mathcal Y}{d\ell}\,\mathbf E_0 ,
\label{eq:curl-h0}
\end{align}
where we used $c\mu\,\mathcal Y=n$; the last term in Eq.~(\ref{eq:curl-h0}) comes from the variation of the
impedance along the ray. Here $\mathbf H_0=\mathcal Y\,\hat{\mathbf p}\times\mathbf E_0$, so that $\rygn\times\mathbf H_0=\hat{\mathbf p}\times d\mathbf H_0/d\ell$, and the derivative of $\mathcal Y$ gives the last term. Since $\hat{\mathbf p}\times\hat{\boldsymbol\nu}=\hat{\boldsymbol\beta}$ and $\hat{\mathbf p}\times\hat{\boldsymbol\beta}=-\hat{\boldsymbol\nu}$, Eq.~(\ref{eq:curl-e0}) gives $\mathbf p\times(\rygn\times\mathbf E_0)=-n\,(D_\nu\hat{\boldsymbol\nu}+D_\beta\hat{\boldsymbol\beta})$, and $-c\mu\,\rygn\times\mathbf H_0=n\,(D_\nu\hat{\boldsymbol\nu}+D_\beta\hat{\boldsymbol\beta})+n\,(d\ln\mathcal Y/d\ell)\,\mathbf E_0$. Hence Equation~(\ref{eq:B-vector}) then yields
\begin{equation}
\boldsymbol{\mathcal C}=n\Big[2\,D_\nu\,\hat{\boldsymbol\nu}+2\,D_\beta\,\hat{\boldsymbol\beta}\Big]
+n\,\frac{d\ln\mathcal Y}{d\ell}\,\mathbf E_0 ,
\label{eq:B-vector-res}
\end{equation}
Setting $\boldsymbol{\mathcal C}=0$ gives Eq.~(\ref{eq:parallel-transport}) of the main text.

\section{Explicit Berry connections and the Frenet-Serret construction}\label{app:frenet}

\subsection{Berry connection and curvature as functions of $\mathbf p$}\label{subsec:Berry-connection}

To obtain $\mathbf A_\sigma(\mathbf p)$ explicitly, choose a constant unit vector $\hat{\mathbf a}$, not collinear with
$\hat{\mathbf p}$, and define
\begin{gather}
\hat{\mathbf e}_1(\mathbf p)=\frac{\hat{\mathbf a}\times\mathbf p}{|\hat{\mathbf a}\times\mathbf p|},
\qquad
\hat{\mathbf e}_2(\mathbf p)=\hat{\mathbf p}\times\hat{\mathbf e}_1(\mathbf p),
\nonumber\\
\hat{\mathbf e}_\sigma(\mathbf p)=\frac{1}{\sqrt2}\big[\hat{\mathbf e}_1+i\sigma\hat{\mathbf e}_2\big] .
\label{eq:e1-e2}
\end{gather}
These are not the Frenet normal and binormal: they depend on $\mathbf p$ only. From
Eq.~(\ref{eq:A_sigma}) and the orthonormality of $\hat{\mathbf e}_1$ and $\hat{\mathbf e}_2$,
$\mathbf A_\sigma=\sigma\,\hat{\mathbf e}_2\cdot\rygnp\hat{\mathbf e}_1$, which gives
\begin{equation}
\mathbf A_\sigma(\mathbf p)=-\sigma\,\frac{(\mathbf p\cdot\hat{\mathbf a})\,[\mathbf p\times\hat{\mathbf a}]}{p\,|\mathbf p\times\hat{\mathbf a}|^2} .
\label{eq:Berry-connection-def}
\end{equation}
For $\hat{\mathbf a}=\hat{\mathbf z}$ and spherical coordinates $(p,\vartheta,\phi)$ of $\mathbf p$, Eq.~(\ref{eq:Berry-connection-def}) is $\mathbf A_\sigma=\sigma\cos\vartheta\,\hat{\boldsymbol\phi}/(p\sin\vartheta)$ and Eq.~(\ref{eq:Berry-connection-def_1}) below is $\tilde{\mathbf A}_\sigma=-\sigma(1-\cos\vartheta)\,\hat{\boldsymbol\phi}/(p\sin\vartheta)$; the radial component of the curl, $(p\sin\vartheta)^{-1}\partial_\vartheta(\sin\vartheta\,A_\phi)$, is $-\sigma/p^2$ for both, and $\mathbf A_\sigma-\tilde{\mathbf A}_\sigma=\sigma\hat{\boldsymbol\phi}/(p\sin\vartheta)=\sigma\rygnp\phi$.  Its curl is the Berry curvature~(\ref{eq:Berry-curvature}), $\rygnp\times\mathbf A_\sigma=-\sigma\mathbf p/p^3$, which does not depend on
$\hat{\mathbf a}$: connections obtained with different vectors $\hat{\mathbf a}$ differ by a gradient
$\rygnp\chi(\mathbf p)$, i.e., by a gauge transformation. An alternative form that is
often used is
\begin{equation}
\tilde{\mathbf A}_\sigma(\mathbf p)=\sigma\,\frac{\mathbf p\times\hat{\mathbf a}}{p\,(p+\mathbf p\cdot\hat{\mathbf a})} ,
\label{eq:Berry-connection-def_1}
\end{equation}
with the same curvature, $\rygnp\times\tilde{\mathbf A}_\sigma=\ryvOm_\sigma$. Explicitly,
$\tilde{\mathbf A}_\sigma=\mathbf A_\sigma-\sigma\rygnp\phi_{\hat a}$, where $\phi_{\hat a}$ is the azimuthal angle of
$\mathbf p$ around $\hat{\mathbf a}$. The connection~(\ref{eq:Berry-connection-def}) is singular
along both half-axes $\mathbf p\parallel\pm\hat{\mathbf a}$, with half-strength strings, and the connection~(\ref{eq:Berry-connection-def_1}) is singular
only along $\mathbf p\parallel-\hat{\mathbf a}$ with a string of full strength. Both connections are perpendicular to $\mathbf p$ and homogeneous of degree
$-1$ in $\mathbf p$.

\subsection{Helicity states in the Frenet basis}

Using Eqs.~(\ref{eq:Frenet-Serret-formulas}) and (\ref{eq:e_sigma}),
\begin{equation}
\frac{d\hat{\mathbf e}_\sigma}{d\ell}=\frac{1}{\sqrt2}\Big[-\mathcal K\,\hat{\mathbf p}+\mathcal T\,\hat{\boldsymbol\beta}-i\sigma\mathcal T\,\hat{\boldsymbol\nu}\Big]
=-i\sigma\,\mathcal T\,\hat{\mathbf e}_\sigma-\frac{\mathcal K}{\sqrt2}\,\hat{\mathbf p} ,
\label{eq:de-frenet}
\end{equation}
because $\mathcal T\hat{\boldsymbol\beta}-i\sigma\mathcal T\hat{\boldsymbol\nu}=-i\sigma\mathcal T(\hat{\boldsymbol\nu}+i\sigma\hat{\boldsymbol\beta})$. The transverse part of
$d\hat{\mathbf e}_\sigma/d\ell$ is proportional to $\hat{\mathbf e}_\sigma$ itself, so that helicity is not changed, and
$\langle\hat{\mathbf e}_\sigma|d\hat{\mathbf e}_\sigma/d\ell\rangle=-i\sigma\mathcal T$. With Eq.~(\ref{eq:A_sigma}), in the Frenet gauge,
\begin{equation}
\mathbf A_\sigma\cdot\frac{d\mathbf p}{d\ell}=i\,\langle\hat{\mathbf e}_\sigma|\tfrac{d}{d\ell}|\hat{\mathbf e}_\sigma\rangle=\sigma\,\mathcal T .
\label{eq:A-frenet}
\end{equation}

\subsection{Relation between the Frenet gauge and the gauge of Eq.~(\ref{eq:e1-e2})}

The Frenet frame $\{\hat{\boldsymbol\nu},\hat{\boldsymbol\beta}\}$ and the frame $\{\hat{\mathbf e}_1,\hat{\mathbf e}_2\}$ of
Eq.~(\ref{eq:e1-e2}) differ by a rotation by the angle $\chi(\ell)$ about $\hat{\mathbf p}$,
$\hat{\boldsymbol\nu}=\cos\chi\,\hat{\mathbf e}_1+\sin\chi\,\hat{\mathbf e}_2$, $\hat{\boldsymbol\beta}=-\sin\chi\,\hat{\mathbf e}_1+\cos\chi\,\hat{\mathbf e}_2$.
Then $\hat{\mathbf e}^{F}_\sigma=e^{-i\sigma\chi}\hat{\mathbf e}_\sigma^{(a)}$, and
\begin{equation}
\sigma\mathcal T=\mathbf A^{(a)}_\sigma\cdot\frac{d\mathbf p}{d\ell}+\sigma\,\frac{d\chi}{d\ell},
\qquad
\int\mathbf A^{(a)}_\sigma\cdot d\mathbf p=\sigma\int\mathcal T\,d\ell-\sigma\,\Delta\chi .
\label{eq:frenet-vs-gauge}
\end{equation}
The Berry phase of Eq.~(\ref{eq:Berry-phase}) in the gauge (a) therefore differs from the torsion phase~(\ref{eq:Berry-phase-frenet}) by the angle $\Delta\chi$ between the two
frames, a gauge term, in agreement with Eq.~(\ref{eq:gauge-phase}).

\section{Centroid of a wave packet}\label{app:centroid}

Consider a wave packet of helicity $\sigma$ in a region where $n$ is approximately constant,
\begin{equation}
\boldsymbol{\mathcal E}(\mathbf r)=\int d^3k\;a(\mathbf k)\,\hat{\mathbf e}_\sigma(\mathbf k)\,e^{i\mathbf k\cdot\mathbf r},
\qquad a(\mathbf k)=|a(\mathbf k)|\,e^{-i\gamma(\mathbf k)} ,
\label{eq:packet}
\end{equation}
with $|a(\mathbf k)|$ concentrated near $\bar{\mathbf k}=k_0\bar{\mathbf p}$. Because the field is transverse, $\hat{\mathbf e}_\sigma(\mathbf k)$ must depend on the direction of $\mathbf k$. Using $\mathbf r\,e^{i\mathbf k\cdot\mathbf r}=-i\rygn_{\mathbf k}e^{i\mathbf k\cdot\mathbf r}$, integrating by parts and using Parseval's theorem, the centroid of $|\boldsymbol{\mathcal E}|^2$ is
\begin{align}
\langle\mathbf r\rangle&=\frac{\int d^3k\,(a\hat{\mathbf e}_\sigma)^*\cdot i\rygn_{\mathbf k}(a\hat{\mathbf e}_\sigma)}{\int d^3k\,|a|^2}
\nonumber\\
&=\frac{\int d^3k\,|a|^2\big[\rygn_{\mathbf k}\gamma+\tilde{\mathbf A}_\sigma(\mathbf k)\big]}{\int d^3k\,|a|^2}
\simeq\rygn_{\mathbf k}\gamma(\bar{\mathbf k})+\frac{1}{k_0}\mathbf A_\sigma(\bar{\mathbf p}) ,
\label{eq:centroid}
\end{align}
where $\tilde{\mathbf A}_\sigma(\mathbf k)=i\hat{\mathbf e}_\sigma^{\,*}\cdot\rygn_{\mathbf k}\hat{\mathbf e}_\sigma=\mathbf A_\sigma(\mathbf p)/k_0$ [Eq.~(\ref{eq:Atilde})], and the term $i|a|\rygn_{\mathbf k}|a|=\tfrac{i}{2}\rygn_{\mathbf k}|a|^2$ integrates to zero. The first term, $\rygn_{\mathbf k}\gamma$, is the stationary-phase point of the scalar phase alone, i.e., the position the packet would have if the polarization vector did not depend on $\mathbf k$; this is the canonical coordinate $\mathbf r$. The second term is the extra displacement caused by the $\mathbf k$ dependence of the polarization vector. Therefore the centroid of the intensity is $\mathbf R=\mathbf r+\mathbf A_\sigma/k_0$. The two terms are separately gauge dependent: the physical field~(\ref{eq:packet}) is unchanged when $\hat{\mathbf e}_\sigma\to e^{i\sigma\chi}\hat{\mathbf e}_\sigma$ and $a\to e^{-i\sigma\chi}a$, so that $\rygn_{\mathbf k}\gamma\to\rygn_{\mathbf k}\gamma+\sigma\rygn_{\mathbf k}\chi$ and $\tilde{\mathbf A}_\sigma\to\tilde{\mathbf A}_\sigma-\sigma\rygn_{\mathbf k}\chi$, while their sum is invariant. Since $\mathbf H_0\propto\hat{\mathbf e}_\sigma$ [Eq.~(\ref{eq:E0-H0-circular})], the magnetic and total energy densities have the same centroid. This is the argument of the semiclassical wave-packet theory \cite{Sundaram1999,Onoda2004,BliokhPRE2004}, and it agrees with the review \cite{Bliokh2009}, where the ray is the trajectory of the centre of gravity of a paraxial beam.

\section{Equivalent formulations of the ray dynamics}\label{app:formulations}

\subsection{Lagrangian formulation in terms of the ray path}\label{subsec:Lagrangian}

Eliminating $\mathbf p$ from the action~(\ref{eq:phase-space-action}) with the zeroth-order relation
\begin{equation}
\mathbf p=n(\mathbf R)\,\frac{\dot{\mathbf R}}{|\dot{\mathbf R}|}+O(k_0^{-1}) ,
\label{eq:p-vs-dot-r-0th-approx}
\end{equation}
where the dot is the derivative with respect to an arbitrary parameter, gives the Lagrangian of the ray path,
\begin{equation}
\mathcal L(\mathbf R,\dot{\mathbf R},\ddot{\mathbf R})=n(\mathbf R)\,|\dot{\mathbf R}|
+\frac{1}{k_0}\,\ddot{\mathbf R}\cdot\mathbf A_\sigma(\dot{\mathbf R}) .
\label{eq:Lagrangian=L_total_final}
\end{equation}
To obtain the second term from $k_0^{-1}\mathbf A_\sigma(\mathbf p)\cdot\dot{\mathbf p}$, note that
$\dot{\mathbf p}=n\ddot{\mathbf R}/|\dot{\mathbf R}|-n(\dot{\mathbf R}\cdot\ddot{\mathbf R})\dot{\mathbf R}/|\dot{\mathbf R}|^3+(\rygn n\cdot\dot{\mathbf R})\dot{\mathbf R}/|\dot{\mathbf R}|$, that the terms along
$\dot{\mathbf R}$ are orthogonal to $\mathbf A_\sigma$, and that $\mathbf A_\sigma$ is homogeneous of degree $-1$,
$\mathbf A_\sigma(\mathbf p)=(|\dot{\mathbf R}|/n)\,\mathbf A_\sigma(\dot{\mathbf R})$. The first term is the usual ray-optics
Lagrangian and the second is the Magnus correction. Because the Lagrangian contains $\ddot{\mathbf R}$, the Euler-Lagrange equation is
$\partial\mathcal L/\partial\mathbf R-\tfrac{d}{d\ell}\,\partial\mathcal L/\partial\dot{\mathbf R}+\tfrac{d^2}{d\ell^2}\,\partial\mathcal L/\partial\ddot{\mathbf R}=0$.
Using $(\mathbf a\cdot\rygn_{\dot R})\mathbf B-\rygn_{\dot R}(\mathbf a\cdot\mathbf B)=-\mathbf a\times(\rygn_{\dot R}\times\mathbf B)$ for constant $\mathbf a$, the terms due to the Berry correction combine to
$-k_0^{-1}\tfrac{d}{d\ell}[\ddot{\mathbf R}\times\ryvOm_\sigma(\dot{\mathbf R})]$, where
$\ryvOm_\sigma(\dot{\mathbf R})=\rygn_{\dot R}\times\mathbf A_\sigma(\dot{\mathbf R})=-\sigma\dot{\mathbf R}/|\dot{\mathbf R}|^3$. The equation becomes
\begin{equation}
\rygn n\,|\dot{\mathbf R}|-\frac{d}{d\ell}\Big(n\frac{\dot{\mathbf R}}{|\dot{\mathbf R}|}\Big)
-\frac{1}{k_0}\frac{d}{d\ell}\Big[\ddot{\mathbf R}\times\ryvOm_\sigma(\dot{\mathbf R})\Big]=0 .
\label{eq:Lagrange_Eqs_motion}
\end{equation}
Choosing the arclength as the parameter after the variation, $|\dot{\mathbf R}|=1$, $\dot{\mathbf R}=\hat{\mathbf t}$, and replacing $\ddot{\mathbf R}$ in the
Berry term with its zeroth-order value, $\ddot{\mathbf R}=\rygn_\perp\ln n$ (the component of $\rygn\ln n$ perpendicular to the ray), we obtain
\begin{equation}
\frac{d}{d\ell}\big(n\,\hat{\mathbf t}\big)=\rygn n+\frac{\sigma}{k_0}\frac{d}{d\ell}\Big[\rygn\ln n\times\hat{\mathbf t}\Big] ,
\label{eq:ray-eq-Magnus}
\end{equation}
where $\rygn_\perp\ln n\times\hat{\mathbf t}=\rygn\ln n\times\hat{\mathbf t}$ was used. The first term is the conventional bending of light toward optically
denser regions, and the second is the Magnus term. Equation~(\ref{eq:ray-eq-Magnus}) is consistent with
Eqs.~(\ref{eq:HamEqr_cor}) and (\ref{eq:HamEqp_cor}): from Eq.~(\ref{eq:HamEqr_cor}), $\hat{\mathbf t}=d\mathbf R/d\ell=\hat{\mathbf p}+\boldsymbol\delta$ with
$\boldsymbol\delta=(\sigma/k_0n^2)\rygn n\times\hat{\mathbf p}$ perpendicular to $\hat{\mathbf p}$, so that $\mathbf p=n(\hat{\mathbf t}-\boldsymbol\delta)$, and
$d\mathbf p/d\ell=\rygn n$ gives $d(n\hat{\mathbf t})/d\ell=\rygn n+d(n\boldsymbol\delta)/d\ell$, which is Eq.~(\ref{eq:ray-eq-Magnus}) to first order in $1/k_0$.

\subsection{Non-canonical bracket form}\label{subsec:brackets}

The ray equations follow from the Hamiltonian $\mathcal H=|\mathbf p|-n(\mathbf R)$ in the variables $(\mathbf R,\mathbf p)$, which obey
non-canonical brackets. From Eq.~(\ref{eq:R-canonical}) and $\{r_i,p_j\}=\delta_{ij}$,
\begin{equation}
\{R_i,p_j\}=\delta_{ij},\qquad
\{R_i,R_j\}=\frac{1}{k_0}\,\varepsilon_{ijl}\,\Omega_{\sigma,l}(\mathbf p) ,
\label{eq:brackets}
\end{equation}
where the second relation follows from $\{R_i,R_j\}=k_0^{-1}(\partial_{p_i}A_{\sigma,j}-\partial_{p_j}A_{\sigma,i})$. The coordinates of the wave
packet do not commute: the Berry curvature plays the role of a magnetic field in momentum space. The equations of motion are
\begin{align}
\frac{dR_i}{d\ell}&=\{R_i,\mathcal H\}=\hat p_i+\{R_i,R_j\}\,\frac{\partial\mathcal H}{\partial R_j},
\\
\frac{dp_i}{d\ell}&=\{p_i,\mathcal H\}=\partial_in(\mathbf R),
\end{align}
which reproduce Eqs.~(\ref{eq:HamEqr_cor}) and (\ref{eq:HamEqp_cor}) because $\partial\mathcal H/\partial R_j=-\partial_jn$. In a quantum treatment,
$[r_i,p_j]=i\delta_{ij}/k_0$ (because $[r_i,k_j]=i\delta_{ij}$ and $\mathbf p=\mathbf k/k_0$), so that the effective Planck constant is $1/k_0$ and Poisson brackets are replaced by commutators
according to $[A,B]=(i/k_0)\{A,B\}$. Then $[R_i,R_j]=i\,\varepsilon_{ijl}\,\Omega_{\sigma,l}/k_0^2=-i\sigma\,\varepsilon_{ijl}\,k_l/k^3$.
\subsection{Frequency as the Hamiltonian}\label{sec:time}

The ray Hamiltonian uses the arclength $\ell$ as the evolution parameter. The other common form takes the frequency as the
Hamiltonian and the time as the evolution parameter,
\begin{equation}
H(\mathbf r,\mathbf k)=\hbar\,\omega(\mathbf r,\mathbf k)=\frac{\hbar c\,|\mathbf k|}{n(\mathbf r)} ,
\label{eq:Homega}
\end{equation}
with canonical momentum $\hbar\mathbf k$. Hamilton's equations are then $\dot{\mathbf r}=\partial\omega/\partial\mathbf k$ and
$\dot{\mathbf k}=-\partial\omega/\partial\mathbf r$, in which the factor $\hbar$ drops out; equivalently, one can take $\omega$ itself as the
Hamiltonian with the bracket $\{r_i,k_j\}=\delta_{ij}$.

\subsubsection{Same constraint surface and reparametrization}\label{subsec:time-reparam}

Fix the frequency $\omega_0=ck_0$ and write $\mathbf k=k_0\mathbf p$. The condition $\omega(\mathbf r,\mathbf k)=\omega_0$ reads
$ck_0|\mathbf p|/n=ck_0$, that is, $|\mathbf p|=n(\mathbf r)$, the eikonal equation~(\ref{eq:eikonal-equation}), which is the constraint
$\mathcal H_0=|\mathbf p|-n=0$ of the ray Hamiltonian without polarization. The two Hamiltonians are related exactly by
\begin{equation}
\omega-\omega_0=\omega_0\Big(\frac{|\mathbf p|}{n}-1\Big)=\frac{\omega_0}{n(\mathbf r)}\,\mathcal H_0(\mathbf r,\mathbf p).
\label{eq:relation}
\end{equation}
In Eq.~(\ref{eq:Homega}) the frequency is the value of the Hamiltonian (an integral of the motion, since $\omega$ has no explicit time dependence),
whereas in the ray Hamiltonian the frequency enters as the fixed parameter $k_0$ and the Hamiltonian takes the value zero.
Equation~(\ref{eq:relation}) shows that the two Hamiltonians differ by a nonvanishing factor $\omega_0/n$ multiplying a function that vanishes on the
physical trajectories. Explicitly, with $\partial/\partial\mathbf k=k_0^{-1}\partial/\partial\mathbf p$ and $\omega=\omega_0|\mathbf p|/n$,
\begin{align}
\frac{d\mathbf r}{dt}&=\frac{\partial\omega}{\partial\mathbf k}=\frac{1}{k_0}\frac{\omega_0}{n}\,\hat{\mathbf p}=\frac{c}{n}\,\hat{\mathbf p},
\label{eq:rdott}\\
\frac{d\mathbf p}{dt}&=-\frac{1}{k_0}\frac{\partial\omega}{\partial\mathbf r}=\frac{c\,|\mathbf p|}{n^{2}}\,\rygn n=\frac{c}{n}\,\rygn n ,
\label{eq:pdott}
\end{align}
where the last step uses $|\mathbf p|=n$. Comparing with Eq.~(\ref{eq:ray0}), these are the equations of Eq.~(\ref{eq:ray0}) multiplied by $c/n$, so that
\begin{equation}
d\ell=\frac{c}{n}\,dt .
\label{eq:dtds}
\end{equation}

\subsubsection{Berry-connection Hamiltonian in the time parametrization}\label{subsec:time-berry}

The shifted-argument Hamiltonian~(\ref{eq:H_tot_corrected}) has a counterpart for Eq.~(\ref{eq:Homega}). Define the Berry connection in $\mathbf k$ space
by
\begin{equation}
\tilde{\mathbf A}_\sigma(\mathbf k)=\frac{1}{k_0}\,\mathbf A_\sigma\!\Big(\frac{\mathbf k}{k_0}\Big),
\qquad
\mathbf F_\sigma(\mathbf k)=\rygn_{\mathbf k}\times\tilde{\mathbf A}_\sigma=-\sigma\,\frac{\mathbf k}{k^{3}} ,
\label{eq:Atilde}
\end{equation}
where the second relation follows from Eq.~(\ref{eq:Berry-curvature}) and $\rygn_{\mathbf k}=k_0^{-1}\rygnp$. The quantity $\tilde{\mathbf A}_\sigma$ has the
dimension of a length, and $\mathbf R=\mathbf r+\tilde{\mathbf A}_\sigma(\mathbf k)$ is the same physical position as in Eq.~(\ref{eq:R-canonical}). The
helicity-dependent Hamiltonian is
\begin{equation}
H_\sigma(\mathbf r,\mathbf k)=\frac{\hbar c\,|\mathbf k|}{n\big(\mathbf r+\tilde{\mathbf A}_\sigma(\mathbf k)\big)} ,
\qquad
\omega_\sigma=\frac{c\,|\mathbf k|}{n(\mathbf R)} .
\label{eq:Hsigma}
\end{equation}
Hamilton's equations are
\begin{align}
\dot r_i&=\frac{\partial\omega_\sigma}{\partial k_i}=\frac{c\,\hat k_i}{n}-\frac{ck}{n^{2}}\,\partial_jn\,\frac{\partial\tilde A_{\sigma,j}}{\partial k_i},\\
\dot k_i&=-\frac{\partial\omega_\sigma}{\partial r_i}=\frac{ck}{n^{2}}\,\partial_in ,
\end{align}
with $n$ and its derivatives evaluated at $\mathbf R$. For the physical position, $\dot R_i=\dot r_i+(\partial\tilde A_{\sigma,i}/\partial k_j)\dot k_j$, and the identity
$\partial_{k_i}\tilde A_{\sigma,j}-\partial_{k_j}\tilde A_{\sigma,i}=\varepsilon_{ijl}F_{\sigma,l}$ [the analogue of Eq.~(\ref{eq:identity})] gives
\begin{equation}
\dot{\mathbf R}=\frac{c}{n}\,\hat{\mathbf k}-\dot{\mathbf k}\times\mathbf F_\sigma(\mathbf k)
=\frac{c}{n}\,\hat{\mathbf k}+\sigma\,\dot{\mathbf k}\times\frac{\mathbf k}{k^{3}},
\qquad
\dot{\mathbf k}=\frac{ck}{n^{2}}\,\rygn n .
\label{eq:Rdott}
\end{equation}
This is the time-parametrized formulation with the Berry curvature $\mathbf F_\sigma$ and the anomalous velocity $-\dot{\mathbf k}\times\mathbf F_\sigma$
\cite{BliokhPRE2004,Bliokh2008,Sundaram1999}. Multiplying Eq.~(\ref{eq:Rdott}) by $dt/d\ell=n/c$ and using $\dot{\mathbf k}=(ck/n^{2})\rygn n$ and $k=k_0n$, the anomalous-velocity term becomes
$\sigma\,(\rygn n\times\hat{\mathbf k})/(nk)=\sigma\,\rygn n\times\hat{\mathbf p}/(k_0n^{2})$, which is the Magnus term~(\ref{eq:Magnus-term}). Thus
Eq.~(\ref{eq:Rdott}) reproduces Eq.~(\ref{eq:HamEqr_cor}) on the surface $p=n$.

\subsubsection{Lagrangian with the Berry term}\label{subsec:time-lagrangian}

The Lagrangian of the Hamiltonian in Eq.~(\ref{eq:Hsigma}) is $L=\mathbf k\cdot\dot{\mathbf r}-\omega_\sigma$. Substituting
$\dot{\mathbf r}=\dot{\mathbf R}-(\dot{\mathbf k}\cdot\rygn_{\mathbf k})\tilde{\mathbf A}_\sigma$ and using
$\mathbf k\cdot(\dot{\mathbf k}\cdot\rygn_{\mathbf k})\tilde{\mathbf A}_\sigma=d(\mathbf k\cdot\tilde{\mathbf A}_\sigma)/dt-\tilde{\mathbf A}_\sigma\cdot\dot{\mathbf k}$ we obtain,
up to a total time derivative,
\begin{equation}
L=\mathbf k\cdot\dot{\mathbf R}+\tilde{\mathbf A}_\sigma(\mathbf k)\cdot\dot{\mathbf k}-\omega(\mathbf R,\mathbf k) .
\label{eq:Lagr-time}
\end{equation}
The term $\tilde{\mathbf A}_\sigma\cdot\dot{\mathbf k}$ is the Berry-phase term, and its variation with respect to $\mathbf k$ gives the anomalous velocity in
Eq.~(\ref{eq:Rdott}). This is the form of the semiclassical Lagrangian of Sundaram and Niu \cite{Sundaram1999}. It is equivalent to the action~(\ref{eq:phase-space-action}):
$\int(\mathbf k\cdot d\mathbf R+\tilde{\mathbf A}_\sigma\cdot d\mathbf k)=k_0\int[\mathbf p\cdot d\mathbf R+k_0^{-1}\mathbf A_\sigma\cdot d\mathbf p]$.

\subsubsection{Summary of the correspondence}\label{subsec:correspondence}

\begin{itemize}
\item Hamiltonian: $\mathcal H_\sigma=|\mathbf p|-n(\mathbf R)$ (ray form) and $H_\sigma/\hbar=ck/n(\mathbf R)$ (time form).
\item Variables: $(\mathbf r,\mathbf p)$ and $(\mathbf r,\mathbf k)$, with $\mathbf k=k_0\mathbf p$.
\item Evolution parameter: $\ell$ and $t$, with $d\ell=(c/n)\,dt$.
\item Shift of the argument of $n$: $\mathbf A_\sigma(\mathbf p)/k_0$ and $\tilde{\mathbf A}_\sigma(\mathbf k)$, the same vector.
\item Curvature: $\ryvOm_\sigma=-\sigma\mathbf p/p^3$ and $\mathbf F_\sigma=-\sigma\mathbf k/k^3$.
\item Frequency: the fixed parameter $k_0$ (the Hamiltonian is zero) and the value of $H/\hbar$.
\end{itemize}

\section{Comparison with Bliokh's review and the helical ray}\label{app:literature}

\subsection{Comparison of conventions and equations}\label{subsec:lit-conventions}

The ray equations~(\ref{eq:HamEqr_cor}) and (\ref{eq:HamEqp_cor}) have exactly the form given in Bliokh's review \cite{Bliokh2009}. There they are obtained from the Coriolis term that appears in the Maxwell equations written in a frame rotating with the ray, and Eq.~(3.40) of the review reads $\dot{\mathbf p}=\rygn n$ and $\dot{\mathbf r}=\mathbf p/p+\bar\lambda_0\,s\,\dot{\mathbf p}\times\mathbf p/p^{3}$, where $\bar\lambda_0=1/k_0$, $s=\pm1$ is the helicity, and the dot is the derivative with respect to the ray length. With $s=\sigma$ and $\dot{\mathbf p}=\rygn n$ this is Eqs.~(\ref{eq:HamEqr_cor}) and (\ref{eq:HamEqp_cor}), including the sign of the Magnus term. The two derivations are therefore consistent. The notation corresponds as follows.
\begin{itemize}
\item The circular polarization vector of the review, $\boldsymbol\xi=(\mathbf v+i\mathbf w)/\sqrt2$, with the right-handed frame $(\mathbf v,\mathbf w,\mathbf t)$,
is the vector $\hat{\mathbf e}_\sigma$ of Eq.~(\ref{eq:e_sigma}) with $\sigma=+1$, with $(\mathbf v,\mathbf w,\mathbf t)\leftrightarrow(\hat{\boldsymbol\nu},\hat{\boldsymbol\beta},\hat{\mathbf p})$
in the Frenet frame.
\item The Berry connection of the review is $\mathbf A^{\rm B}=-i\,\boldsymbol\xi^{*}\cdot\rygnp\boldsymbol\xi$ (its Eq.~(3.17)) and the curvature is
$\mathbf F=\mathbf p/p^{3}$ [Eq.~(3.20)], with the helicity entering as an explicit factor $s$. In this paper the helicity is inside the connection,
$\mathbf A_\sigma=-\sigma\mathbf A^{\rm B}$ and $\ryvOm_\sigma=-\sigma\mathbf F$.
\item The spin-orbit Lagrangian $-\bar\lambda_0\,s\,\mathbf A^{\rm B}\cdot\dot{\mathbf p}$ of the review [Eq.~(3.39)] is the term $+k_0^{-1}\mathbf A_\sigma\cdot d\mathbf p/d\ell$ of
the action~(\ref{eq:phase-space-action}), and varying $\mathbf r$ and $\mathbf p$ independently gives Eqs.~(\ref{eq:HamEqr_cor}) and (\ref{eq:HamEqp_cor}), as in the review. The
coordinate $\mathbf r$ of the review is the physical position, our $\mathbf R$.
\item The phase of the review is $\Phi=\Phi_0-s\Phi_{\rm B}$ with $\Phi_{\rm B}=\int\mathbf A^{\rm B}\cdot d\mathbf p=-\int\mathcal T\,d\ell$ in the Frenet frame, which
is the phase $\varphi_\sigma=\int\mathbf A_\sigma\cdot d\mathbf p=\sigma\int\mathcal T\,d\ell$ of Eqs.~(\ref{eq:Berry-phase}) and (\ref{eq:Berry-phase-frenet}). The review also gives the
parallel-transport law for the polarization vector, $\dot{\mathbf e}=-(\mathbf e\cdot\dot{\mathbf t})\mathbf t$ [Eq.~(3.32)], which is our Eq.~(\ref{eq:parallel-transport}) for a constant impedance.
\item The Hamiltonian of the review is $\mathcal H=p-n-\bar\lambda_0\,s\,\Omega_\parallel$ [Eq.~(3.11)], where $\Omega_\parallel$ is the component along $\mathbf t$ of the angular
velocity of the ray frame, which is the torsion in the Frenet frame. The interaction term of the expanded Hamiltonian~(\ref{eq:H_tot_corrected}),
$-k_0^{-1}\mathbf A_\sigma\cdot d\mathbf p/d\ell$, equals $-(\sigma/k_0)\mathcal T$ in the Frenet gauge by Eq.~(\ref{eq:A-frenet}), which is the same term.
\end{itemize}
Two differences should be noted. The review assumes a nonmagnetic medium ($n^2=\epsilon/\epsilon_0$), whereas here $\epsilon(\mathbf r)$ and $\mu(\mathbf r)$ both vary. Within the
approximations of Sec.~\ref{sec:maxwell}, the parallel-transport law, and hence the Berry phase and the ray equations, depend only on $n$; the impedance enters only through the
amplitude factor ${\mathcal Z}^{1/2}$ of Eq.~(\ref{eq:E_sigma-solution}). Second, the review describes the rays as the trajectories of the center of gravity of a paraxial wave packet or beam,
which supports the identification of $\mathbf R$ with the center of the wave packet used in Sec.~\ref{subsec:Hamiltonian}\ and derived in Appendix~\ref{app:centroid}.

\subsection{Helical ray in a cylindrically symmetric medium}\label{subsec:lit-analytic}

For $n=n(\rho)$ in cylindrical coordinates $(\rho,\Phi,Z)$, the scalar ray equations have helical solutions of radius $\rho$ with tangent
$\hat{\mathbf p}=\sin\theta\,\mathbf e_\Phi+\cos\theta\,\mathbf e_Z$, $\mathbf p=n\hat{\mathbf p}$, where
\begin{equation}
\sin^{2}\theta=-\frac{\rho\,n'(\rho)}{n(\rho)}
\end{equation}
(Eq.~(4.3) of the review \cite{Bliokh2009}). The curvature and the torsion of the helix are $\mathcal K=\sin^2\theta/\rho$ and $\mathcal T=\sin\theta\cos\theta/\rho$, and the Frenet vectors are
$\hat{\boldsymbol\nu}=-\mathbf e_\rho$ and $\hat{\boldsymbol\beta}=-\cos\theta\,\mathbf e_\Phi+\sin\theta\,\mathbf e_Z$. With $\rygn n=n'\mathbf e_\rho$ one finds from Eq.~(\ref{eq:Magnus-binormal})
\begin{equation}  \label{eq:dRMdell}
\frac{d\mathbf R_{\rm M}}{d\ell}=-\sigma\,\frac{\sin^{2}\theta}{k_0\,n\,\rho}\,\hat{\boldsymbol\beta} ,
\end{equation}
which is Eq.~(4.13) of the review \cite{Bliokh2009} (the instantaneous Magnus velocity).  

Integrating the velocity~in Eq.~(\ref{eq:dRMdell}) at fixed $\mathbf p$ over one period of the helix, $\ell=2\pi\rho/\sin\theta$ (the assumption $\delta\mathbf p=0$ of the review, its Eq.~(3.45)), gives $-\sigma\lambda\sin\theta\,\hat{\boldsymbol\beta}$ with $\lambda=2\pi/[k_0n(\rho)]$, the value of Eq.~(4.15) of the review. This is the anomalous-velocity part only. A helix whose momentum is unchanged but whose tangent is rotated by the Magnus term does not satisfy $d\mathbf p/d\ell=\rygn n(\mathbf R)$ of Eq.~(\ref{eq:HamEqp_cor}): with $\mathbf R'=\hat{\mathbf p}-\sigma\mathcal K\hat{\boldsymbol\beta}/(k_0n)$ the azimuth advances at the rate $\Phi'=[\sin\theta_p+\sigma\mathcal K\cos\theta_p/(k_0n)]/\rho$, whereas the radial force requires $\Phi'=-n'/(n\sin\theta_p)$. Equating the two gives the momentum angle $\theta_p$ of the exact helix of radius $\rho$,
\begin{equation}
\sin^2\theta_p=\sin^2\theta-\sigma\,\frac{\sin^2\theta}{k_0\,n\,\rho}\,\sin\theta_p\cos\theta_p
\ \Rightarrow\ \theta_p\simeq\theta-\frac{\sigma\mathcal K}{2k_0n},
\end{equation}
while the tangent of the ray is at the angle $\theta+\sigma\mathcal K/(2k_0n)$, so that the ray is rotated, relative to the $\sigma=0$ helix, by half of the angle $\sigma\mathcal K/(k_0n)$ that is obtained at fixed $\mathbf p$. After one period the displacement is therefore
\begin{equation}
\delta\mathbf R_0=-\sigma\,\frac{\lambda}{2}\,\sin\theta\;\hat{\boldsymbol\beta},\qquad\lambda=\frac{2\pi}{k_0n(\rho)},
\label{eq:helix-shift}
\end{equation}
and the splitting between $\sigma=+1$ and $\sigma=-1$ is $\lambda\sin\theta$ per period. We checked this by integrating Eqs.~(\ref{eq:num-R}) and (\ref{eq:num-p}) numerically for $n(\rho)=n_a(1-\rho^2/2\Lambda^2)$ with $\rho/\Lambda=0.5$ and $k_0=400$ (lengths in units of $\Lambda$): started on the exact helix, the ratio of the computed displacement to Eq.~(\ref{eq:helix-shift}) is $1.000$ and to the value of the review $0.500$, in the same direction. If both rays are started with the same $\mathbf p_0$, the displacement after one period is $0.62$ of the value of the review in this example, because an oscillation of the radius of the same order is superposed on the drift.
The torsion phase per period is
$\sigma\int\mathcal T\,d\ell=2\pi\sigma\cos\theta$; modulo $2\pi$ it equals $-\sigma\Theta$, where $\Theta=2\pi(1-\cos\theta)$ is the solid angle enclosed on the sphere of directions, in agreement with the
statement after Eq.~(\ref{eq:gauge-phase}) and with Eq.~(4.11) of the review. The shift accumulates linearly with the number of periods.


\begin{thebibliography}{99}
\bibitem{J-example}
For example, in the gauge~(\ref{eq:Berry-connection-def_1}), using $(\mathbf p\times\hat{\mathbf a})\times\mathbf p=p^2\hat{\mathbf a}-(\mathbf p\cdot\hat{\mathbf a})\mathbf p$,
\begin{equation}
\mathbf J=\mathbf r\times\mathbf p+\frac{\sigma}{k_0}\,\frac{\hat{\mathbf a}+\hat{\mathbf p}}{1+\hat{\mathbf p}\cdot\hat{\mathbf a}} ,
\label{eq:J-canonical-gauge}
\end{equation}
which is singular only on the Dirac string $\hat{\mathbf p}=-\hat{\mathbf a}$. Its component along $\hat{\mathbf a}$ is $(\mathbf r\times\mathbf p)\cdot\hat{\mathbf a}+\sigma/k_0$. Hence, if $\hat{\mathbf a}$ is chosen along a symmetry axis of the medium, the canonical orbital angular momentum about that axis is itself conserved, and it differs from $\mathbf J\cdot\hat{\mathbf a}$ only by the constant $\sigma/k_0$.
\bibitem{classical-commutator}
The same result follows from the symmetry of the problem. With the non-canonical brackets, Eq.~(\ref{eq:brackets}), $\{R_i,R_j\}=k_0^{-1}\epsilon_{ijl}\Omega_{\sigma,l}$, $\{R_i,p_j\}=\delta_{ij}$, $\{p_i,p_j\}=0$, the orbital part alone does not generate rotations of $\mathbf R$. One finds $\{(\mathbf R\times\mathbf p)_i,R_j\}=\epsilon_{ijl}R_l+k_0^{-1}(p_j\Omega_{\sigma,i}-\delta_{ij}\,\mathbf p\cdot\ryvOm_\sigma)$, and the extra term is cancelled exactly by $\{(\sigma/k_0)\hat p_i,R_j\}=-(\sigma/k_0)(\delta_{ij}-\hat p_i\hat p_j)/p$ for $\ryvOm_\sigma=-\sigma\mathbf p/p^3$. Therefore
\begin{equation*}
\{J_i,R_j\}=\epsilon_{ijl}R_l ,\qquad \{J_i,p_j\}=\epsilon_{ijl}p_l ,
\end{equation*}
so $\mathbf J$ is the generator of rotations of the ray variables $(\mathbf R,\mathbf p)$. By Noether's theorem $d\mathbf J/d\ell=\{\mathbf J,\mathcal H\}$ vanishes whenever the Hamiltonian $\mathcal H=|\mathbf p|-n(\mathbf R)$ is rotationally invariant. The spin term plays the same role as the Poincar\'e term in the angular momentum of a charge in the field of a magnetic monopole; here the monopole (the Berry curvature) is in momentum space.
\end{thebibliography}
\end{document}